\documentclass{cernyrep}
\usepackage[T1]{fontenc}
\usepackage[bookmarks, colorlinks=true, linktoc=page, pdftex, linkcolor=black, citecolor=black, urlcolor=blue]{hyperref}

\usepackage{fancyhdr}
\fancyhfoffset{4 mm}
\fancypagestyle{ARTTITLE}{%
\fancyhf{} 
\lhead{\hspace*{0cm} CERN Accelerator School Proceedings ---
{\it Beam Instrumentation} ---  Split,  Croatia, 2025\hfill}
\lfoot{\hspace{3mm} Available online at \url{https://cas.web.cern.ch/previous-schools}}
\rfoot{\thepage\hspace*{3mm}}
 
}

\usepackage{varwidth}
\usepackage{xcolor}
\begin{document}

\title{ Digital Signal Processing for Beam Instrumentation}
\author{ Andrea Boccardi}
\institute{CERN, Geneva, Switzerland}

\begin{abstract}
These proceedings summarise a lecture on Digital Signal Processing (DSP) delivered at the CERN Accelerator School (CAS) on Beam Instrumentation (BI).

In this context, a signal is any measurable quantity conveying information about the beam. A digital signal is obtained by discretising a continuous-time signal in both time (sampling) and amplitude (quantisation). Digital signal processing refers to the set of techniques used to analyse and manipulate such signals.

Given the breadth of the field, this contribution focuses on fundamental DSP concepts and on techniques most commonly encountered in beam instrumentation applications.
\end{abstract}
\maketitle
\thispagestyle{ARTTITLE}

\section{Introduction}

Digital signal processing is an essential element of modern beam instrumentation, where precise measurement and analysis of signals are essential for monitoring and controlling particle beams. In this context, a signal conveys information about the beam, such as position, intensity, or frequency characteristics. A digital signal is obtained by sampling a continuous signal in time and quantising it in amplitude, enabling its processing by computational methods.

The objective of this lecture is to provide a foundation in the fundamental concepts and techniques of DSP as applied to beam instrumentation. While DSP encompasses a vast field, this contribution focuses on the basic techniques most relevant to accelerator diagnostics. These include sampling, filtering, and spectral analysis, which are routinely used in the acquisition and processing of beam signals. By emphasising practical examples and commonly encountered scenarios, the lecture illustrates how digital signal processing affects the precision, reliability, and flexibility of beam instrumentation systems.

The foundation of the DSP techniques discussed in this lecture is frequency-domain analysis, originating from Fourier series theory. In the first section, we review the essential concepts, starting from the Fourier series (FS), through the Fourier transform (FT) and the Discrete-Time Fourier Transform (DTFT), and progressing to the Discrete Fourier Transform (DFT), while highlighting key properties and theorems.

This discussion is not intended to provide a fully rigorous mathematical derivation, but rather to emphasise the key concepts and their connections in an intuitive way. Readers interested in formal derivations or a more detailed treatment of the presented formulas and theorems are encouraged to consult dedicated textbooks \cite{bib:OppenheimSS,bib:OppenheimDSP,bib:Lyons}.

We then discuss filters in the digital domain, with a brief introduction to the Z transform. The~discussion emphasises their applications in beam instrumentation, such as noise reduction, signal conditioning, and frequency selection.

These concepts are then illustrated through three complete case studies drawn from real CERN beam instrumentation systems: single-bunch position measurement in the CERN SPS, the choice between a matched filter and an energy estimator for the HL-LHC beam position monitor system, and IQ demodulation in the CERN LEIR orbit system.

The final part of the text focuses on the numerical representation of numbers, from integer to floating-point formats, as well as on the different hardware solutions available for DSP implementation in beam instrumentation.

\section{Recap of Frequency Domain Analysis}

\subsection{The Fourier Series and the Fourier Transform}

The Fourier series (FS) represents a periodic signal as an infinite sum of trigonometric functions (sines and cosines) whose frequencies are integer multiples of the fundamental frequency of the signal \cite{bib:OppenheimSS}.

For a periodic signal \(s_p(t)\) of period \(T = 1/f_0\), the \textbf{\textit{Fourier series}} can be expressed as:

\begin{align}
s_p(t) &= \frac{a[0]}{2} + \sum_{k=1}^{\infty} \Big[ a[k] \cos(2 \pi k f_0 t) + b[k] \sin(2 \pi k f_0 t) \Big] \label{eq:fourier_series} \\
a[k]  &= \frac{2}{T} \int_{-\frac{T}{2}}^{\frac{T}{2}} s(t) \cos(2 \pi k f_0 t) \, dt \\
b[k]  &= \frac{2}{T} \int_{-\frac{T}{2}}^{\frac{T}{2}} s(t) \sin(2 \pi k f_0 t) \, dt
\end{align}

Alternatively, the Fourier series can be expressed in complex exponential form:

\begin{align}
s_p(t) &= \sum_{n=-\infty}^{\infty} c[n]  e^{j 2 \pi n f_0 t} \label{eq:fourier_series_complex} \\
c[k]  &= \frac{1}{T} \int_{-\frac{T}{2}}^{\frac{T}{2}} s_p(t)  e^{-j 2 \pi k f_0 t} dt \label{eq:fourier_series_complex_analysis}
\end{align}

The complex coefficients \(c[n]\) can be expressed in terms of their \emph{magnitude} and \emph{phase}. These quantities can be plotted as functions of frequency, generating the magnitude spectrum and the phase spectrum of the signal.

Since the Fourier series describes a periodic signal, these spectra are discrete and exist only at harmonic frequencies \(f[k] = k f_0\) where \(f_0 = 1/T\) is the fundamental frequency of the signal.

\begin{figure}[!ht]
    \centering
    \includegraphics[width=1.0 \textwidth]{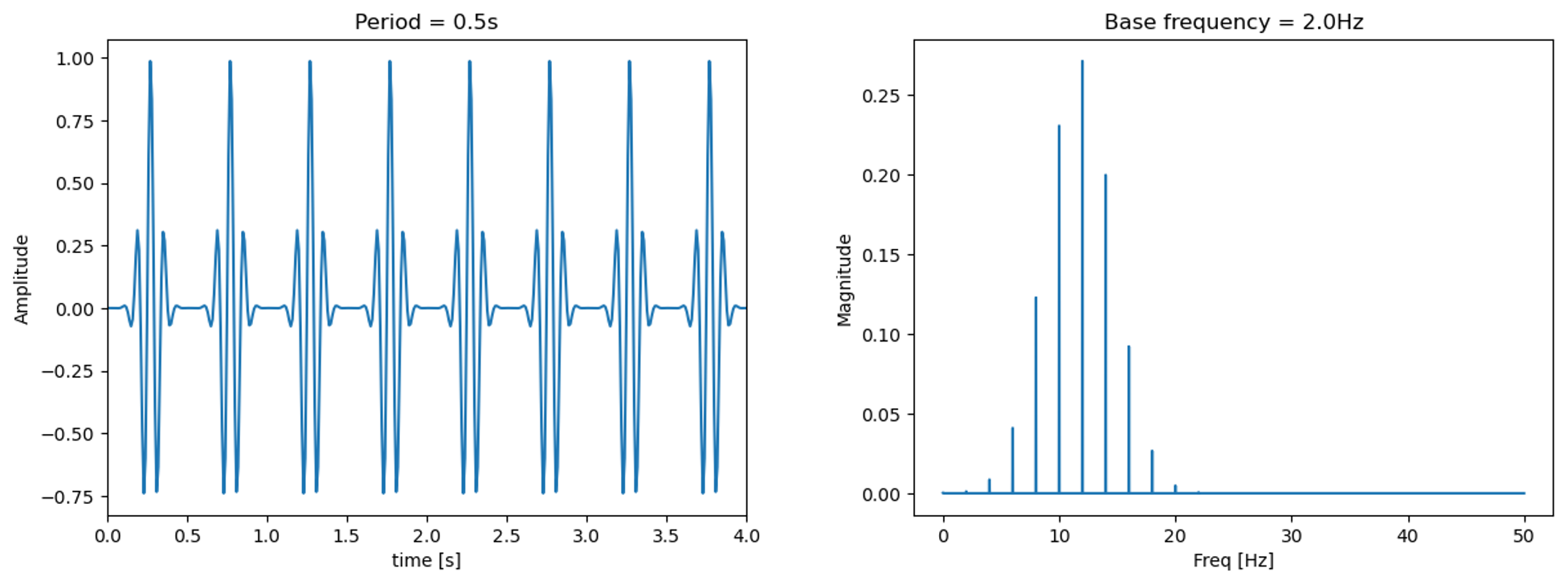}
    \caption{Example of a periodic signal (left) and its magnitude spectrum (right). The spectrum consists of discrete lines located at harmonic frequencies \(f[k] = k f_0\). Only the positive frequency axis is displayed.}
    \label{fig:fs_spectrum_example}
\end{figure}

An example of a periodic signal and its corresponding magnitude spectrum is illustrated in Fig.~\ref{fig:fs_spectrum_example} (only positive frequencies are shown for clarity).

The lower the fundamental frequency \(f_0\), the denser the spectrum. As a conceptual exercise, consider increasing the period of the signal \(s_p(t)\) to \(T' = mT\), with \(m \in \mathbb{N}\), by inserting intervals of zero value in between each basic period \(T\). 

For the harmonics of \(f_0\), the integral in Eq.~(\ref{eq:fourier_series_complex_analysis}) extended over the new period differs only by the~scaling factor \(1/m\). At the same time, the spectral lines become increasingly dense, with \(m-1\) additional harmonics appearing between every pair of original harmonics.

In the limit \(m \rightarrow +\infty\), the spacing between adjacent harmonics tends to zero. The discrete coefficients \(c[n]\) evolve into a continuous function \(S(f)\), and the summation becomes an integral. In this limit, the time domain function can be seen as not periodic with \(s_p(t) \rightarrow s(t)\). This leads naturally to the~Fourier transform, which applies to non-periodic signals \cite{bib:OppenheimSS}. The synthesis
and analysis formulas of the~\textbf{\textit{Fourier transform}} are given by:

\begin{align}
s(t) = \int_{-\infty}^{\infty} S(f) e^{j 2\pi f t} df \label{eq:fourier_transform_synthesis} \\
S(f) = \int_{-\infty}^{\infty} s(t) e^{-j 2\pi f t} dt \label{eq:fourier_transform_analysis}
\end{align}

This in the following text will be expressed as: \(s(t)
\quad \underleftrightarrow{\quad \mathcal{F} \quad} \quad
S(f)\).

\begin{figure}[!ht]
    \centering
    \includegraphics[width=1.0\textwidth]{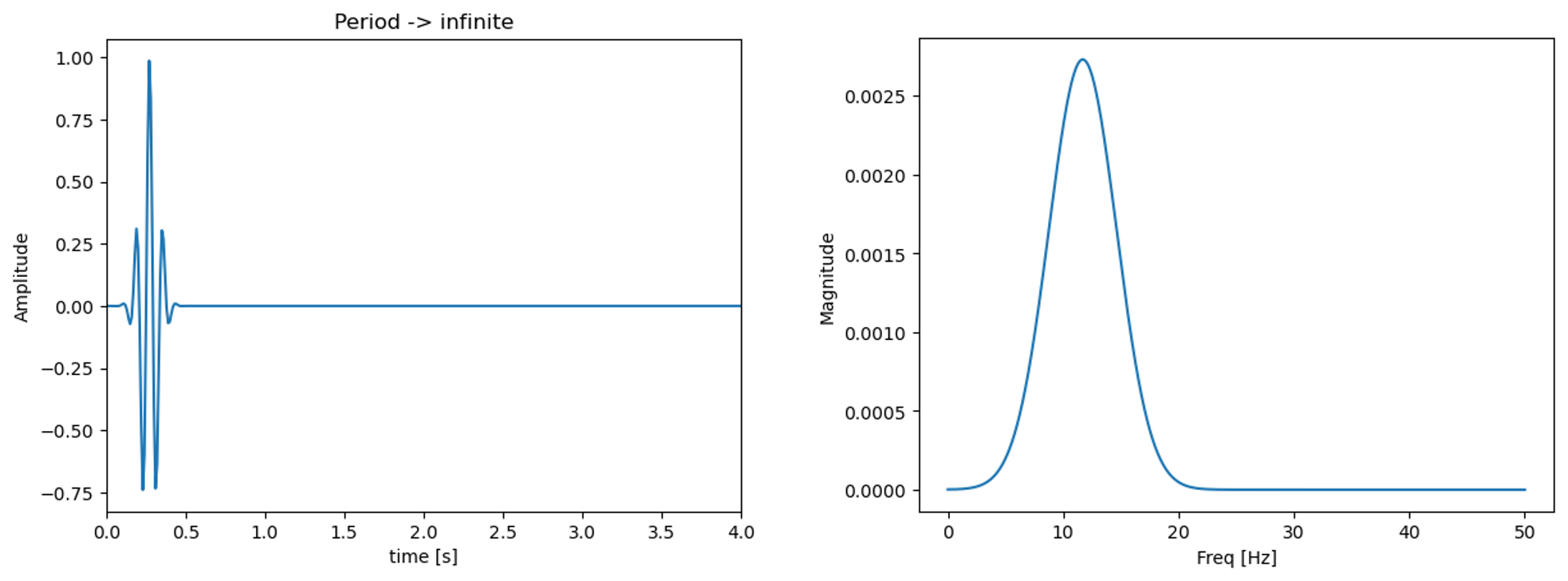}
    \caption{Signal obtained by isolating one base period of the signal shown in Fig.\ref{fig:fs_spectrum_example} (left), and its corresponding continuous magnitude spectrum given by the Fourier transform (right).}
    \label{fig:ft_extension}
\end{figure}

\subsubsection{Useful Fourier transforms}

It is useful to recall some notable Fourier transforms which will be used later in the text.

\paragraph{Cosine}
The Fourier transform of a cosine (or a sine), simply consists of 2 Dirac delta, \(\delta(\cdot)\), at opposite frequencies. For a cosine signal of unit amplitude and frequency \(f_c\),

\begin{equation}
\cos(2\pi f_c t) 
\quad \underleftrightarrow{\quad \mathcal{F} \quad} \quad 
\frac{1}{2} \left[ \delta(f - f_c) + \delta(f + f_c) \right].
\end{equation}

\paragraph{Dirac delta} 
The Dirac delta function has a flat spectrum over all frequencies:

\begin{equation}
\delta(t) 
\quad \underleftrightarrow{\quad \mathcal{F} \quad} \quad 
1
\end{equation}

\paragraph{Dirac comb} 
The Dirac comb (or impulse train), defined as a series of Dirac deltas 
spaced by intervals \(T\), has as Fourier transform another series of Dirac deltas, now spaced in 
frequency by \(1/T\):

\begin{equation}
\sum_{n=-\infty}^{\infty} \delta(t - nT)
\quad \underleftrightarrow{\quad \mathcal{F} \quad} \quad 
\frac{1}{T} \sum_{k=-\infty}^{\infty} 
\delta\!\left(f - \frac{k}{T}\right)
\end{equation}

\paragraph{Rectangular pulse} 
The rect is a very relevant function in DSP and is often used to isolate finite portions of a signal, as will be shown later. For the normalised rectangular function we have:

\begin{equation}
\mathrm{rect}(t) =
\begin{cases}
1, & |t| < \tfrac{1}{2} \\
0, & |t| > \tfrac{1}{2} \\
\tfrac{1}{2}, & |t| = \tfrac{1}{2}
\end{cases}
\quad \underleftrightarrow{\quad \mathcal{F} \quad} \quad
\mathrm{sinc}(f) = \frac{\sin(\pi f)}{\pi f}.
\end{equation}

\paragraph{Gaussian function} 
A Gaussian function maintains its functional shape under Fourier transformation, but the \(\sigma\) in the de-numerator of the exponent in the time domain appears in the numerator of the exponent in the frequency domain. This makes particularly evident a fundamental time–frequency 
duality: the wider the function in time (larger \(\sigma\)), 
the narrower its spectrum in frequency, and vice versa. For a normalised Gaussian with standard deviation \(\sigma\) we have:

\begin{equation}
\frac{1}{\sqrt{2\pi\sigma^2}} 
e^{-\frac{t^2}{2\sigma^2}}
\quad \underleftrightarrow{\quad \mathcal{F} \quad} \quad
e^{-2\pi^2\sigma^2 f^2}
\end{equation}

\subsubsection{Relevant properties of the Fourier transform}

Key properties that will be used extensively include linearity, duality, time scaling, symmetry for real signals, time shift, differentiation, and convolution. These properties form the basis for understanding sampling and filtering operations \cite{bib:OppenheimSS,bib:OppenheimDSP}.

\paragraph{Linearity}
The Fourier transform is a linear operator. If
\begin{equation*}
s_a(t) \quad \underleftrightarrow{\quad \mathcal{F} \quad} \quad S_a(f), 
\quad
s_b(t) \quad \underleftrightarrow{\quad \mathcal{F} \quad} \quad S_b(f),
\end{equation*}
then for any constants \(a, b \in \mathbb{C}\),
\begin{equation}
a s_a(t) + b s_b(t) 
\quad \underleftrightarrow{\quad \mathcal{F} \quad} \quad
a S_a(f) + b S_b(f).
\end{equation}

\paragraph{Duality}
If
\begin{equation*}
s(t) \quad \underleftrightarrow{\quad \mathcal{F} \quad} \quad S(f),
\end{equation*}
then, by duality,
\begin{equation}
S(t) \quad \underleftrightarrow{\quad \mathcal{F} \quad} \quad s(-f).
\end{equation}
This property highlights the symmetry between the time and frequency domains.

\paragraph{Time scaling}
If
\begin{equation*}
s(t) \quad \underleftrightarrow{\quad \mathcal{F} \quad} \quad S(f),
\end{equation*}
then
\begin{equation}
s(at) 
\quad \underleftrightarrow{\quad \mathcal{F} \quad} \quad
\frac{1}{|a|} S\!\left(\frac{f}{a}\right),
\quad a \neq 0.
\end{equation}

Thus, compression in time corresponds to expansion in frequency, and vice versa.

\paragraph{Symmetry for real signals}
If \(s_R(t)\) is real-valued, then its spectrum satisfies the Hermitian symmetry property:
\begin{equation}
s_R(t) \in \mathbb{R},\quad S_R(-f) = S_R^*(f)
\end{equation}
where \(^*\) denotes complex conjugation. 
As a consequence, the magnitude spectrum is even and the phase spectrum is odd.

\paragraph{Time shift (delay)}
A time delay introduces a linear phase term in the frequency domain:
\begin{equation}
s(t - t_0)
\quad \underleftrightarrow{\quad \mathcal{F} \quad} \quad
S(f) e^{-j 2\pi f t_0}.
\end{equation}

\paragraph{Time differentiation}
Differentiation in time corresponds to multiplication by \(j 2\pi f\) in frequency:
\begin{equation}
\frac{d}{dt} s(t)
\quad \underleftrightarrow{\quad \mathcal{F} \quad} \quad
j 2\pi f \, S(f).
\end{equation}

\paragraph{Convolution}
The convolution of two signals \(s(t)\) and \(h(t)\) is defined as
\begin{equation*}
s(t) \otimes h(t) = \int_{-\infty}^{\infty} s(\tau)\, h(t - \tau)\, d\tau.
\end{equation*}

Where \(\otimes\) is the convolution operator. The Fourier transform converts convolution in time into multiplication in frequency:
\begin{equation}
s(t) \otimes h(t)
\quad \underleftrightarrow{\quad \mathcal{F} \quad} \quad
S(f) H(f).
\end{equation}

Conversely, multiplication in time corresponds to convolution in frequency:
\begin{equation}
s(t) h(t)
\quad \underleftrightarrow{\quad \mathcal{F} \quad} \quad
S(f) \otimes H(f).
\end{equation}

This result is particularly relevant for the sampling and for filtering.

\paragraph{Parseval's theorem (Fourier series)}
For a periodic signal \(s(t)\) of period \(T\), expressed in terms of its 
complex Fourier Series coefficients

\[
s_p(t) = \sum_{k=-\infty}^{\infty} c[k
]e^{j 2\pi k f_0 t},
\quad f_0 = \frac{1}{T},
\]

Parseval's theorem states that the average power of the signal over one period 
is equal to the sum of the squared magnitudes of its Fourier series coefficients:

\begin{equation}
\frac{1}{T} \int_{0}^{T} |s_p(t)|^2 \, dt
=
\sum_{n=-\infty}^{\infty} |c[n]|^2.
\end{equation}

\subsection{From continuous to discrete time and frequency domains}

\subsubsection{Sampling and Shannon--Nyquist sampling theorem}
Time-domain sampling of a signal can be modelled as the multiplication of a continuous-time signal by a~Dirac comb, where the impulses are spaced by the sampling period \(T\). The sampled signal can be written as:

\begin{equation}
s_s(t) = s(t)\sum_{n=-\infty}^{\infty} \delta(t - nT_s).
\end{equation}

In the frequency domain, by virtue of the convolution theorem, this operation corresponds to the~convolution of the signal spectrum with a Dirac comb in frequency. The resulting spectrum is a~periodic replication of the original spectrum at frequencies \(f_n = n f_s\), where \(f_s = 1/T_s\) is the sampling frequency:

\begin{equation}
S_s(f) = \frac{1}{T_s} \sum_{n=-\infty}^{\infty} S(f - n f_s).
\label{eq:sampling_replicas}
\end{equation}

If the original signal is band-limited with bandwidth \(B\), i.e.\ all its energy is contained within the~frequency range \([-B, B]\), and if the sampling frequency satisfies the \textbf{\textit{Nyquist condition}} \(f_s \geq 2B\), then the spectral replicas are fully separated.

Under this condition, the signal in the \textbf{\textit{base-band}}, defined as \(\left[-\frac{f_s}{2}, \frac{f_s}{2}\right]\), is an exact copy of the~original spectrum. This implies that no information is lost during the sampling process. This result is known as the Shannon--Nyquist sampling theorem \cite{bib:Shannon,bib:OppenheimDSP}.

If the Nyquist condition is not satisfied, sampling introduces aliasing, i.e.\ an overlap of the spectral replicas, which results in distortion of the reconstructed signal.

In the specific case of a cosine signal, aliasing manifests as an apparent folding of the true frequency into the base-band, so that different original frequencies become indistinguishable after sampling.

\subsubsection{The Discrete Time Fourier Transform (DTFT)}
If we use the expression of \(s_s(t)\) to compute its Fourier Transform (see Eq.~(\ref{eq:fourier_transform_analysis})), since the sampled signal consists of weighted Dirac impulses located at the discrete time instants \(t = nT_s\), the Fourier-transform integral reduces to a summation over the sample sequence.

The synthesis equation of the Fourier transform (see Eq.~(\ref{eq:fourier_transform_synthesis})) remains an integral over frequency, but it can be restricted to a fundamental interval as we assume the Nyquist condition to hold.

In particular, defining the discrete-time signal \(s[n] = s(nT_s)\) as the sequence of samples of the~original signal, we obtain a representation of the sampled signal in terms of a sequence indexed by \(n\), and the synthesis and analysis formulas of this discrete-time signal are given by:

\begin{equation}
s[n] = \int_{-f_s/2}^{f_s/2} S(f)\, e^{j2\pi fnT_s}\, df
\end{equation}

\begin{equation}
S(f) = \sum_{n=-\infty}^{\infty} s[n]\, e^{-j2\pi f nT_s}
\end{equation}

This is known as the Discrete Time Fourier Transform (DTFT), which, by defining the normalised dimensionless frequency \(\Omega = f/f_s\), can be expressed as:

\begin{equation}
s[n] = \int_{-1/2}^{1/2} S(\Omega)\, e^{j2\pi \Omega n}\, d\Omega.
\end{equation}

\begin{equation}
S(\Omega) = \sum_{n=-\infty}^{\infty} s[n]\, e^{-j2\pi \Omega n}
\label{eq:DTFT_analysis}
\end{equation}

The DTFT is periodic in \(\Omega\) with period 1, which reflects the inherent periodicity introduced by time-domain sampling \cite{bib:OppenheimDSP}.

\subsubsection{Discrete Fourier Transform (DFT)}
The implementation of the DTFT in digital logic or software presents two main limitations when applied to acquired signals. First, the summation in Eq.~(\ref{eq:DTFT_analysis}) extends from \(-\infty\) to \(+\infty\), and second, the frequency variable is continuous.

While one could in principle evaluate the transform at a finite set of frequencies of interest, thereby addressing the issue of the continuous frequency axis, the requirement of observing the signal over an~infinite time interval remains a fundamental limitation, unless the analytical expression of the signal is known.

\paragraph{Windowing}
In practical applications, signals acquired from sensors are observed over a finite time interval multiple of the sampling period \(T_s\), referred to as the observation window \(T_o=N T_s\), with \(N \in \mathbb{N}\). The signal is therefore assumed to be zero outside this interval. As a result, the DTFT summation is effectively computed over the \(N\) acquired samples.

This operation is known as \emph{windowing}. Mathematically, it corresponds to multiplying the original signal by a time-limited function, like the rectangular pulse:

\begin{equation}
s_w[n] = s[n]\, \mathrm{rect}\!\left[\frac{n}{N}\right].
\end{equation}

\begin{figure}[!ht]
    \centering
    \includegraphics[width=1.0\textwidth]{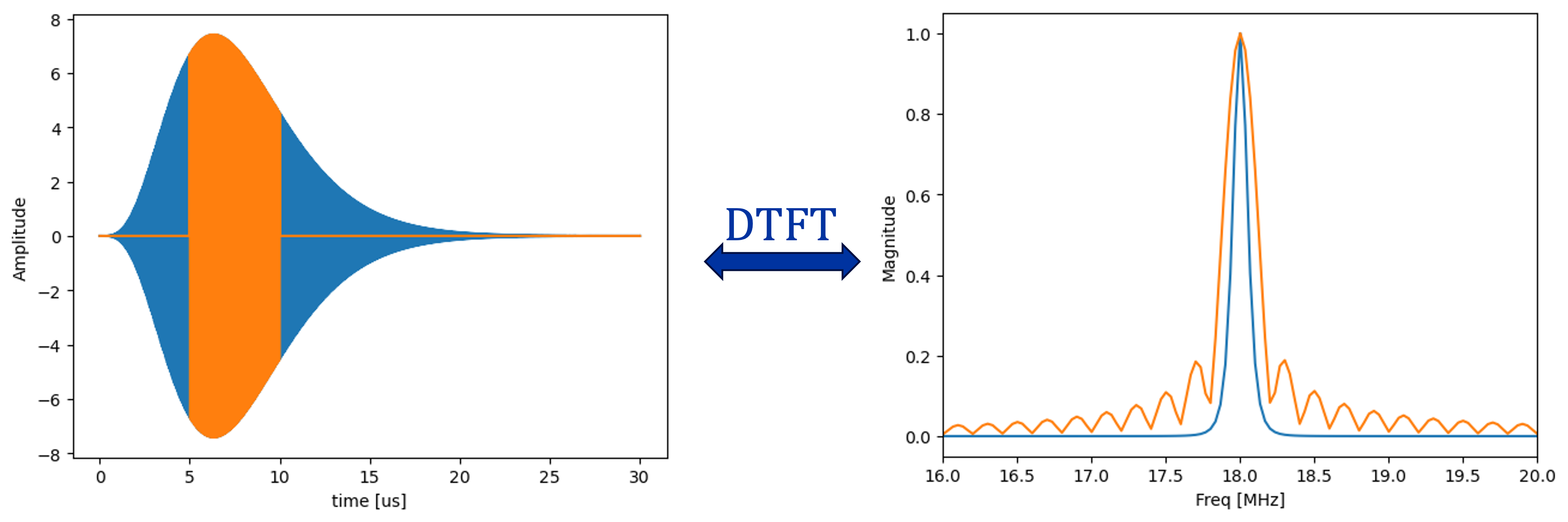}
    \caption{Example of the effect of applying a rectangular window. On the left, the time-domain signal is shown: in blue the original signal, and in orange its windowed version. On the right, the corresponding magnitude spectra are displayed. In this example, the acquisition window is relatively short with respect to the signal bandwidth. The~resulting convolution with the \(sinc\) function in the frequency domain leads to a widening of the main lobe and the~appearance of side lobes (spectral leakage).}
    \label{fig:windowing_effect}
\end{figure}

In the frequency domain, this multiplication results in a convolution between the original spectrum and the spectrum of the window function. Consequently, the observed spectrum is distorted, reflecting the modification of the signal in the time domain (see Figure~\ref{fig:windowing_effect}).

The duration of the window plays a key role: the longer the observation time, i.e. the larger the~number of samples \(N\), the narrower the corresponding transform of the window, and therefore the~smaller the spectral distortion.

Beyond the rectangular pulse, different window functions can be used, each providing a trade-off between spectral resolution and amplitude accuracy; the Hann and the Hamming being two among the~most commonly used.

The rectangular window corresponds to a simple truncation of the signal in time and, among the~windowing functions mentioned, provides the best frequency resolution, as its main lobe in the~frequency domain is the narrowest. However, it also exhibits the highest side lobes, which results in significant spectral leakage and can mask weaker frequency components.

The Hann window reduces spectral leakage by smoothly tapering the signal to zero at the edges of the observation interval. This significantly lowers the side lobes in the frequency domain, at the cost of a~wider main lobe, and therefore reduced frequency resolution. 
\begin{equation}
w_{Hann}[n] =
\begin{cases}
\dfrac{1}{2} \left(1 - \cos\left(\dfrac{2\pi n}{N-1}\right)\right), & 0 \leq n \leq N-1, \\[1ex]
0, & \text{otherwise}.
\end{cases}
\end{equation}

Similarly, the Hamming window provides a compromise between the rectangular and Hann windows. It offers improved side-lobe suppression compared to the rectangular window, while maintaining a slightly narrower main lobe than the Hann window. 
\begin{equation}
w_{Hamming}[n] =
\begin{cases}
0.54 - 0.46 \cos\left(\dfrac{2\pi n}{N-1}\right), & 0 \leq n \leq N-1, \\[1ex]
0, & \text{otherwise}.
\end{cases}
\end{equation}

Those are just three examples of the many windowing functions that can be used \cite{bib:Harris}. 

More generally, window selection depends on the application: if resolving closely spaced frequency components is critical, a narrow main lobe is preferred, whereas if detecting weak signals in the~presence of strong components is required, lower side lobes are more advantageous.

In beam instrumentation applications, this trade-off is particularly relevant, as signals often contain components with very different amplitudes and closely spaced frequencies.

In the remainder of this text we will assume the use of the rectangular window.

\paragraph{Periodicisation of the time-limited signal}
While windowing makes the DTFT applicable to practical signals, one additional step is required to arrive at the form commonly implemented in digital signal processing. The finite-duration signal can be considered as the base period of a periodic signal with period \(T_p \geq T_o\). This is equivalent to convolving the windowed signal with a Dirac comb of period \(T_p\) in the time domain.

\begin{figure}[!ht]
    \centering
    \includegraphics[width=1.0\textwidth]{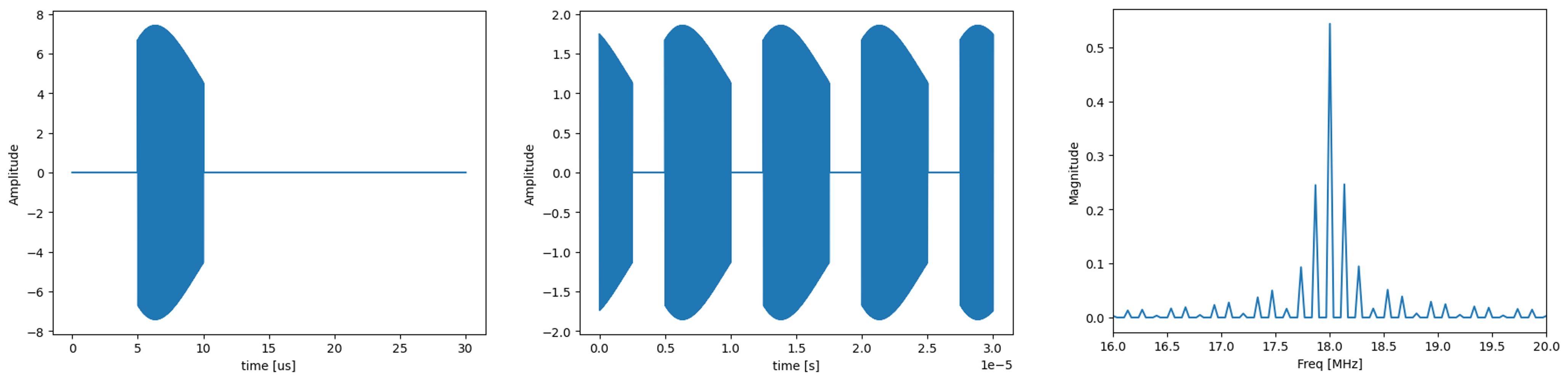}
    \caption{In this figure, the periodicisation process is illustrated: the final step that leads from the DTFT to the DFT. The plot on the left represents the windowed signal from Fig.~\ref{fig:windowing_effect}, the centre plot represents its periodicised version, and the one on the right shows the resulting magnitude spectrum. It should be noted that the period \(T_p = N_p T_s\) of the resulting signal has been chosen to be longer than \(N T_s\), where N is the number of acquired samples, in order to make the periodic structure in time more clearly visible in the plot. Extending the base period also has the effect of providing a finer sampling of the frequency-domain representation, and is sometimes used for this purpose.}
    \label{fig:periodisation_effect}
\end{figure}

In the frequency domain, this corresponds to multiplying the spectrum by a Dirac comb with spacing \(1/T_p\), effectively sampling the continuous frequency axis at discrete frequencies \(f_k=\frac{k}{T_p}\). It is common to have \(N_p=N\), and this is what it will be assumed in the remainder of this proceedings.

The resulting transform is discrete in frequency and is known as the \textbf{\textit{Discrete Fourier Transform (DFT)}}:

\begin{equation}
s_w[n] = \frac{1}{N} \sum_{k=0}^{N-1} S_w[k]\, e^{j2\pi \frac{k}{N} n}
\end{equation}

\begin{equation}
S_w[k] = \sum_{n=0}^{N-1} s_w[n]\, e^{-j2\pi \frac{k}{N} n}
\end{equation}

\paragraph{Notes on the Fast Fourier Transform (FFT)}
If \(N\) is a power of 2, particularly efficient algorithms can be used to compute the transform, significantly reducing the number of required operations from \(\mathcal{O}(N^2)\) to \(\mathcal{O}(N \log_2 N)\). These algorithms are commonly known as the Fast Fourier Transform (FFT) \cite{bib:CooleyTukey}.

When the first DFT algorithms were implemented in software libraries, the hardware limitations of the time made computational efficiency a critical requirement. As a consequence, the transform was almost always implemented using FFT algorithms in order to reduce memory usage and computational time.

This often favoured input arrays whose length was a power of two. In some cases, this constraint was enforced internally by extending the input array with additional zeros, resulting in an output spectrum with a different number of samples than the original input signal.

Modern FFT libraries efficiently support arbitrary transform lengths, although powers of two often remain slightly advantageous from a computational perspective.

The term FFT is still commonly used to refer to the computation of the DFT, independently of the~specific implementation algorithm.

\begin{figure}[!ht]
    \centering
    \includegraphics[width=1.0\textwidth]{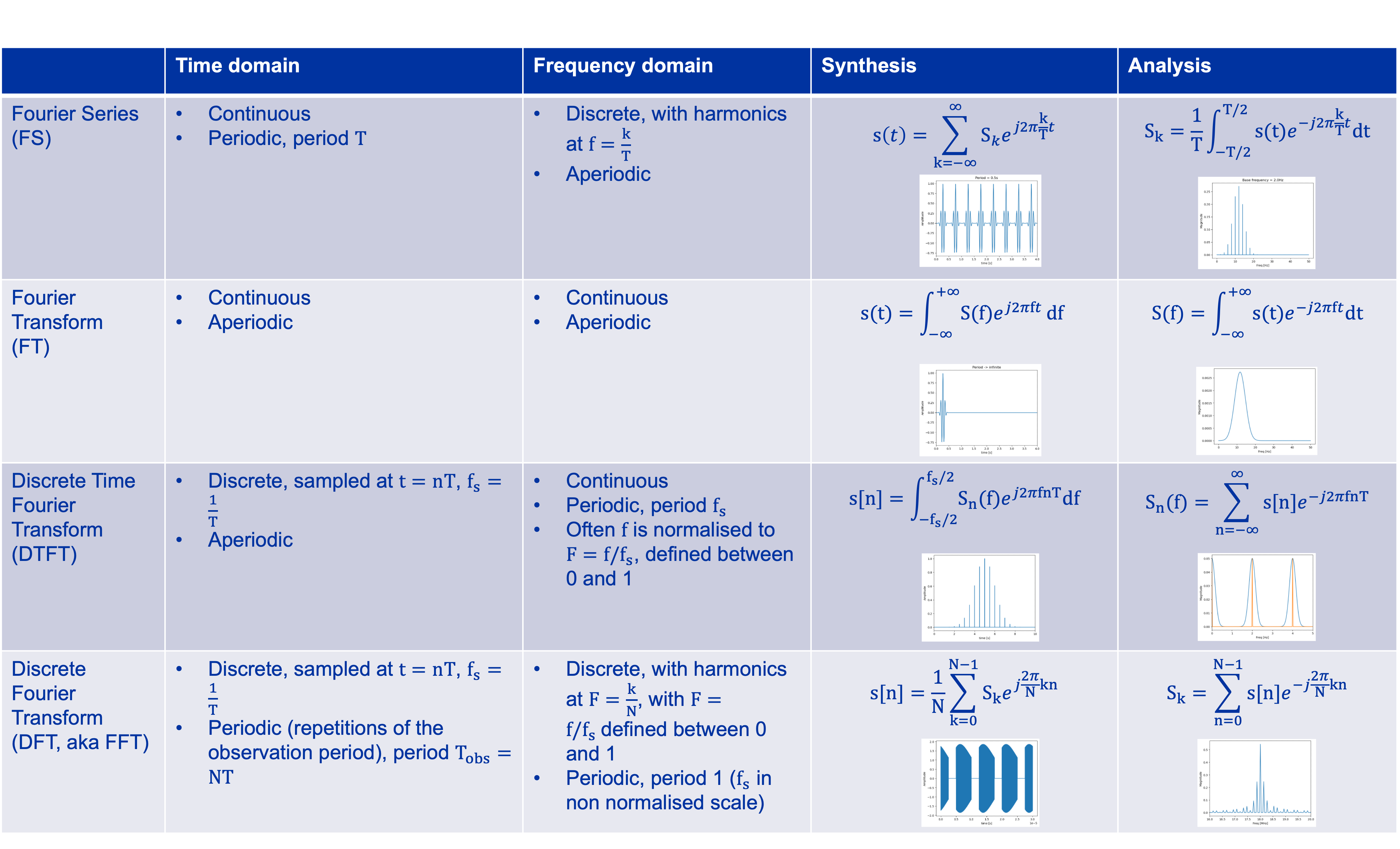}
    \caption{Summary of the basic properties and formulas of the different Transforms.}
    \label{fig:transforms_summary}
\end{figure}

\section{Filters}

\subsection{Basic concepts}

In common language, a filter is something that selects certain features and excludes others. In signal processing, filters are most often used to select or exclude specific frequency ranges -- for instance to exclude the frequency bands where there is no power from the signal of interest but only noise, or to reject a specific, unwanted frequency that couples into the measurement without being part of the signal itself.

The most common types of filters encountered in practice are:

\begin{itemize}
\item \textbf{Low-pass (LP) filters}: let the lower-frequency part of the spectrum pass, up to a frequency called the \emph{cut-off} frequency, and attenuate the rest;
\item \textbf{High-pass (HP) filters}: let the higher-frequency part of the spectrum pass, above a cut-off frequency, and attenuate the rest;
\item \textbf{Band-pass (BP) filters}: let a band of frequencies pass, defined by a low and a high cut-off frequency, and attenuate the rest;
\item \textbf{Notch filters}: strongly attenuate one specific frequency and let all the rest pass;
\item \textbf{Resonant filters}: let a very narrow band of frequencies pass, defined by a centre frequency and a~bandwidth.
\end{itemize}

Regardless of the specific filter type, it is useful to define generally the \textbf{\textit{pass-band}} as the range of frequencies that a filter is designed to let through with little or no attenuation, and the \textbf{\textit{stop-band}} as the range of frequencies that it is designed to substantially attenuate or reject. For a low-pass filter, for instance, the pass-band lies below the cut-off frequency and the stop-band above it; for a band-pass filter, the pass-band lies between the two cut-off frequencies and the stop-band outside them. Between the~pass-band and the stop-band lies the \textbf{\textit{transition band}}, where the filter response falls off gradually rather than switching abruptly from one regime to the other.

One of the key parameters of LP, HP and BP filters is their cut-off frequency (or frequencies). This is defined as the frequency at which the power of the signal is attenuated by a factor of 2 with respect to the pass-band level. This point is commonly referred to as the \(-3\)~dB point, with dB standing for \emph{decibel}.

The decibel is a logarithmic unit used to express the ratio between two power levels \(P\) and \(P_{\mathrm{ref}}\),

\begin{equation}
\left[\mathrm{dB}\right] = 10\, \log_{10}\!\left(\frac{P}{P_{\mathrm{ref}}}\right),
\end{equation}

or, equivalently, since power is proportional to the square of the signal amplitude \(A\),

\begin{equation}
\left[\mathrm{dB}\right] = 20\, \log_{10}\!\left(\frac{A}{A_{\mathrm{ref}}}\right).
\end{equation}

From this definition, a power ratio of \(1/2\) corresponds to \(10\log_{10}(1/2) \approx -3.01\)~dB, which is the origin of the \(-3\)~dB cut-off convention introduced above.

For LP, HP and BP filters, the cut-off frequency is probably the main defining characteristic, but it is far from the only relevant one; on its own, it would give a very limited understanding of the properties and quality of a filter. Other important characteristics include:

\begin{itemize}
\item the \textbf{roll-off}, i.e.\ the steepness of the transition between pass-band and stop-band, typically expressed in dB per decade (or per octave) of frequency, and directly related to the order (number of poles) of the filter;
\item the \textbf{pass-band ripple}, i.e.\ how flat the response is within the pass-band -- some filter families deliberately trade a small amount of ripple for a sharper transition;
\item the \textbf{stop-band attenuation}, i.e.\ how strongly the unwanted frequencies are suppressed;
\item the \textbf{phase response}, or equivalently the group delay, describing how much the filter delays different frequency components of the signal -- a non-linear phase response distorts the shape of the filtered signal even when the magnitude response is close to ideal;
\item for resonant and notch filters specifically, the \textbf{quality factor} \(Q\), defined as the ratio of the centre (or notch) frequency to the \(-3\)~dB bandwidth of the resonance or notch, \(Q = f_{\mathrm{centre}}/\Delta f_{-3\mathrm{dB}}\): the higher the \(Q\), the narrower and sharper the resonance or notch.
\end{itemize}

The remainder of this section examines filters in more detail, and how to analyse and design them.

\subsection{The impulse response of a filter}

A filter can be thought of as selecting parts of the spectrum of a signal by directly multiplying it by a~mask function in the frequency domain, \(Y(f) = H(f)\,X(f)\), where \(X(f)\) is the spectrum of the input signal and \(H(f)\) the filter mask, or \textbf{\textit{transfer function}}. By the convolution theorem already introduced in Section~2, this is equivalent, in the time domain, to the convolution of the input signal \(x(t)\) with \(h(t)\), the~inverse Fourier transform of \(H(f)\):

\begin{equation}
y(t) = h(t) \otimes x(t).
\end{equation}

The function \(h(t)\) is known as the \textbf{\textit{impulse response}} of the filter, since it is precisely the output of the filter when excited with a Dirac delta, \(\delta(t)\): recalling that \(\delta(t) \underleftrightarrow{\ \mathcal{F}\ } 1\), the response to a unit impulse is \(h(t) \underleftrightarrow{\ \mathcal{F}\ } H(f)\cdot 1 = H(f)\), i.e.\ the impulse response and the transfer function form a Fourier pair.

Filter responses are commonly visualised using \textbf{\textit{Bode diagrams}}: a pair of plots showing the magnitude of \(H(f)\), expressed in dB, and its phase, both as a function of frequency on a logarithmic scale. Representing the magnitude in dB and the frequency axis logarithmically is particularly convenient because, as will become clear later in this section once poles and zeros have been introduced, the contribution of each individual pole and zero of a filter becomes a simple straight-line asymptote on such a diagram, and the response of several filter stages in cascade can be obtained simply by adding their individual Bode plots.

\subsection{Degression on analogue filters}

While this is a course on digital signal processing, it is instructive to first look at filters in the analogue domain. Analogue filters are commonly used to prototype and design digital ones, since a much larger body of literature and design methods exists for them, and it is generally easier, mathematically, to design the desired behaviour starting from their equations and only later convert the result to the digital domain, rather than starting directly in the digital domain.

A filter, like any signal-processing system, is said to be \textbf{\textit{linear}} if it satisfies the superposition principle -- i.e.\ its response to a weighted sum of inputs equals the same weighted sum of the individual responses -- and \textbf{\textit{time-invariant}} if a time shift of the input simply produces an identical time shift of the output, with no other change. A system satisfying both properties is referred to as a \textbf{\textit{linear time-invariant (LTI)}} system; all the filters considered in this text, both in the analogue and in the digital domain, belong to this class.

In the analogue domain, LTI filters can be described by a differential equation with constant coefficients, relating the output \(y(t)\) to the input \(x(t)\):

\begin{equation}
\sum_{i=0}^{M} a_i \frac{d^i}{dt^i} y(t) = \sum_{i=0}^{N} b_i \frac{d^i}{dt^i} x(t).
\label{eq:analogue_diff_eq}
\end{equation}

The linearity of this equation follows from it being a weighted sum of the input, output and their derivatives, with no higher-order or cross terms; time-invariance, in turn, is precisely why the coefficients \(a_i\), \(b_i\) are constants, independent of time -- were they instead functions of \(t\), a time-shifted input would no longer generally produce a simply time-shifted output.

Recalling the differentiation property of the Fourier transform (\(\frac{d}{dt}s(t) \underleftrightarrow{\ \mathcal{F}\ } j2\pi f\, S(f)\)), Eq.~(\ref{eq:analogue_diff_eq}) can be rewritten in the frequency domain as

\begin{equation}
\sum_{i=0}^{M} a_i (j2\pi f)^i\, Y(f) = \sum_{i=0}^{N} b_i (j2\pi f)^i\, X(f),
\end{equation}

which leads to the transfer function of the filter, expressed either as a ratio of sums or, after factorising numerator and denominator, as a ratio of products:

\begin{equation}
H(f) = \frac{Y(f)}{X(f)} = \frac{\sum_{i=0}^{N} b_i (j2\pi f)^i}{\sum_{i=0}^{M} a_i (j2\pi f)^i} = K\, \frac{\prod_{i=0}^{N} \left(1 - j\frac{f}{f_{0i}}\right)}{\prod_{i=0}^{M} \left(1 - j\frac{f}{f_{pi}}\right)}.
\label{eq:analogue_transfer_function}
\end{equation}

The \(f_{0i}\) are called the \textbf{\textit{zeros}} and the \(f_{pi}\) the \textbf{\textit{poles}} of the filter. Focusing on the magnitude of the~response, which determines which frequencies are attenuated and which are preserved, Eq.~(\ref{eq:analogue_transfer_function}) shows that it is straightforward to decompose the magnitude response into the individual contribution of each zero and pole:

\begin{equation}
20\log_{10}|H(f)| = 20\log_{10}|K| + \sum_{i=0}^{N} 20\log_{10}\left|1-j\frac{f}{f_{0i}}\right| - \sum_{i=0}^{M} 20\log_{10}\left|1-j\frac{f}{f_{pi}}\right|,
\label{eq:magnitude_decomposition}
\end{equation}

i.e.\ on a Bode diagram, the total magnitude response (in dB) is simply the sum of the individual contributions of each zero (added) and each pole (subtracted).

For the contribution of a single pole, \(1/\left|1-jf/f_{p}\right|\), two asymptotic straight lines can readily be identified: for \(f \ll |f_p|\), the value can be approximated by 1 (0~dB, a flat line), while for \(f \gg |f_p|\), the function can be approximated by a straight line of slope \(-1\) in log-log coordinates -- equivalent to \(-20\)~dB per decade on the Bode magnitude diagram -- passing through the point \((f_p, 1)\).

\begin{figure}[!ht]
    \centering
    \includegraphics[width=1.0\textwidth]{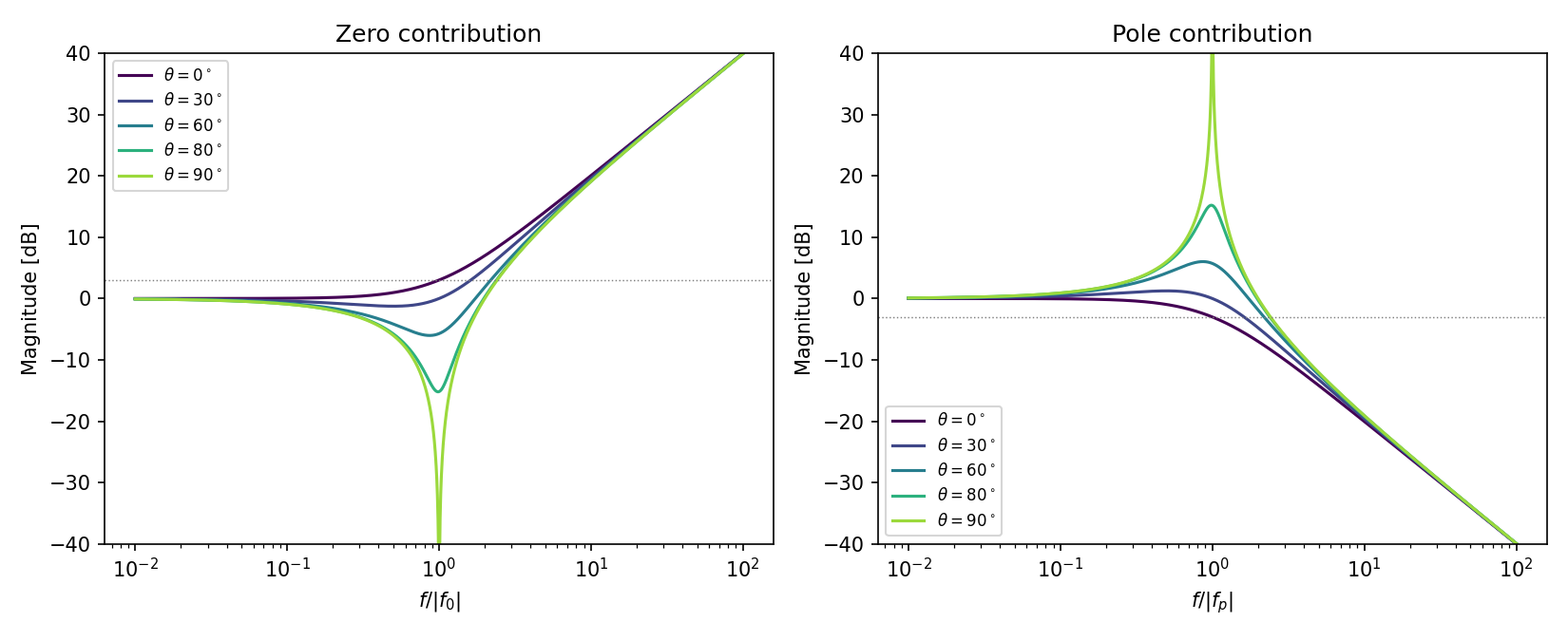}
    \caption{Magnitude contribution, in dB, of a single zero (left) and a single pole (right) located at \(f_0=|f_0|e^{j\theta}\) (respectively \(f_p\)), as a function of \(f/|f_0|\) (respectively \(f/|f_p|\)), for several values of the angle \(\theta\) between \(0^\circ\) (purely real, smooth response) and \(90^\circ\) (purely imaginary, sharp notch or resonance). The dotted horizontal line marks the~\(\pm 3\)~dB level reached at \(f=f_0\) (resp.\ \(f_p\)) in the purely real case.}
    \label{fig:bode_zero_pole_family}
\end{figure}

The exact shape of the curve between these two asymptotes depends on the ratio between the real and imaginary parts of \(f_p\), which is in general a complex number. For \(f_p\) purely real, the curve is smooth and monotonic, with a value of \(1/\sqrt{2}\) (i.e.\ \(-3\)~dB) at \(f=f_p\) -- the same \(-3\)~dB convention introduced for the cut-off frequency in Section~3.1, here appearing directly as the familiar single-pole low-pass roll-off. For \(f_p\) purely imaginary, on the other hand, a much sharper feature appears: the magnitude diverges to \(+\infty\)~dB at \(f=f_p\), the signature of an ideal, lossless resonance. All other cases lie between these two extremes, and the sharpness of the feature is governed by the ratio between the imaginary and real parts of \(f_p\) -- directly related to the quality factor \(Q\) introduced in Section~3.1, which increases as \(f_p\) approaches the purely imaginary case. This family of curves is shown in the right panel of Fig.~\ref{fig:bode_zero_pole_family}.

The same considerations apply to the contribution of a zero, \(\left|1-jf/f_{0}\right|\), except that, being in the~numerator of \(H(f)\) rather than the denominator, its effect on the total response is inverted: a purely real zero produces a smooth, monotonically rising curve, \(+3\)~dB at \(f=f_0\) -- here appearing as a bump rather than a dip because \(f_0\) is a zero rather than a pole -- while a purely imaginary zero produces a sharp dip, dropping exactly to zero (\(-\infty\)~dB) at \(f=f_0\), the signature of an ideal notch. This is illustrated in the~left panel of Fig.~\ref{fig:bode_zero_pole_family}.

\begin{figure}[!ht]
    \centering
    \includegraphics[width=1.0\textwidth]{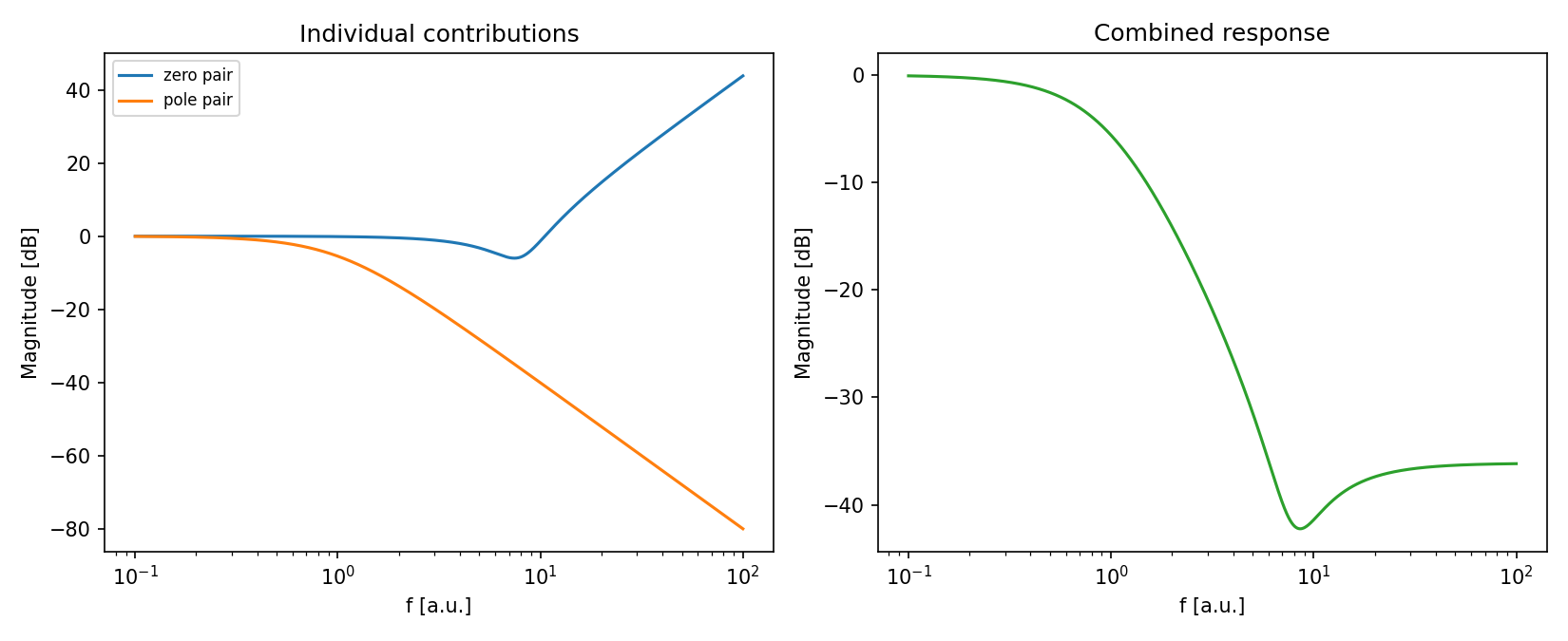}
    \caption{Example of combining a pole pair (producing a low-pass roll-off) and a zero pair (producing a shallow notch-like dip) to build a more complex filter response: individual contributions (left) and combined magnitude response (right).}
    \label{fig:bode_combination}
\end{figure}

In general, poles and zeros are combined to produce more complex frequency-selective behaviours. As an example, Fig.~\ref{fig:bode_combination} shows the individual contributions of a zero pair (placed with a large imaginary component, producing a shallow, comparatively sharp dip) and a pole pair (placed with a large real component, producing a smooth low-pass roll-off), together with the resulting combined response -- a low-pass filter with an additional, superimposed dip at the zero-pair frequency.

\section{Digital filters}

\subsection{Basic concepts}

The mathematical formalism for digital filters is, in the time domain, actually simpler than for their analogue counterparts: for an LTI filter, it reduces to a linear combination of the current and past input samples (\emph{feedforward} terms) and of past output samples (\emph{feedback} terms),

\begin{equation}
y[n] = \sum_{i=0}^{N-1} b_i\, x[n-i] + \sum_{i=1}^{M-1} a_i\, y[n-i].
\label{eq:general_difference_equation}
\end{equation}

Recalling the time-shift property of the DTFT, \(x[n-i] \underleftrightarrow{\ \ } X(\Omega)\, e^{-j2\pi\Omega i}\), the DTFT of each term of Eq.~(\ref{eq:general_difference_equation}) can be taken directly, giving

\begin{equation}
Y(\Omega) = \sum_{i=0}^{N-1} b_i\, X(\Omega)\, e^{-j2\pi\Omega i} + \sum_{i=1}^{M-1} a_i\, Y(\Omega)\, e^{-j2\pi\Omega i}.
\end{equation}

For convenience, it is useful to define the complex variable \(Z=e^{j2\pi\Omega}\) and, correspondingly, \(Z^{-1}\) as the \textbf{\textit{delay operator}},

\begin{equation}
x[n-i] \ \underleftrightarrow{\ \ }\ Z^{-i}X(Z),
\end{equation}

so that the equation above can be rewritten simply as

\begin{equation}
Y(Z) = \sum_{i=0}^{N-1} b_i\, Z^{-i}\, X(Z) + \sum_{i=1}^{M-1} a_i\, Z^{-i}\, Y(Z).
\end{equation}

At this point, it is straightforward to write the transfer function of the filter as a function of \(Z\):

\begin{equation}
H(Z) \triangleq \frac{Y(Z)}{X(Z)} = \frac{\sum_{i=0}^{N-1} b_i\, Z^{-i}}{1 - \sum_{i=1}^{M-1} a_i\, Z^{-i}} = b_0\, \frac{\prod_{i=1}^{N-1} \left(1 - Z_{0i}Z^{-1}\right)}{\prod_{i=1}^{M-1} \left(1 - Z_{pi}Z^{-1}\right)},
\label{eq:HZ_general}
\end{equation}

where \(b_0\) is simply the coefficient of the \(i=0\) tap already appearing in the sum above, and \(Z_{0i}\), \(Z_{pi}\) are, respectively, the zeros and the poles of the filter -- the digital-domain equivalent of the \(f_{0i}\), \(f_{pi}\) of the analogue transfer function, Eq.~(\ref{eq:analogue_transfer_function}). Note that, unlike the analogue gain \(K\), \(b_0\) is generally \emph{not} equal to the digital filter's DC gain \(H(1)\).

\begin{figure}[!ht]
    \centering
    \includegraphics[width=0.5\textwidth]{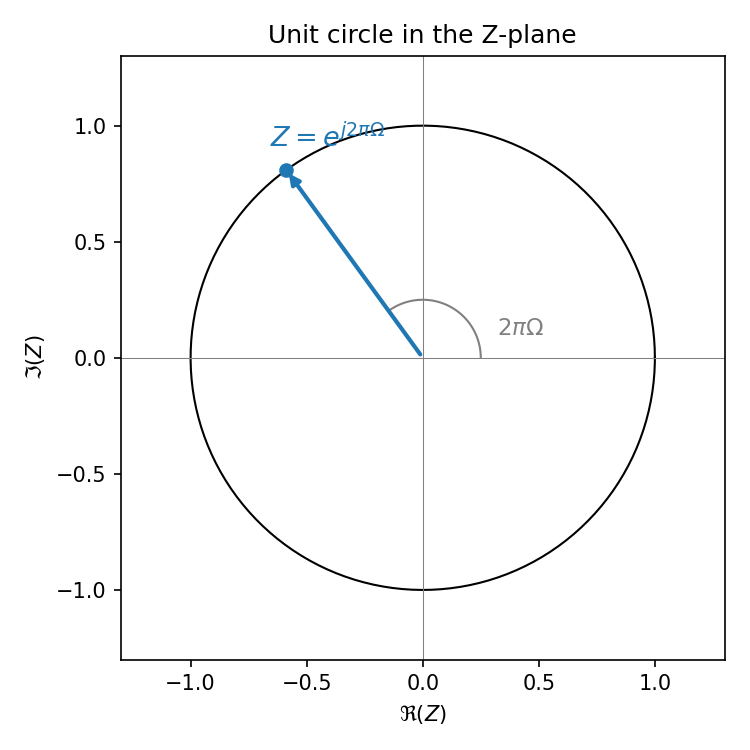}
    \caption{The unit circle in the \(Z\)-plane. A given normalised frequency \(\Omega\) corresponds to the point \(Z=e^{j2\pi\Omega}\) on the~circle, at an angle \(2\pi\Omega\) from the positive real axis.}
    \label{fig:z_unit_circle}
\end{figure}

This illustrates how easy it is, in the digital domain, to go from the time-domain equation of a~filter to its representation in the frequency domain, simply by introducing the variable \(Z\). In doing so, we have just encountered the \textbf{\textit{Z-transform}}, introduced here in a deliberately simplified way -- a full treatment is beyond the scope of this lecture, but it is important to know that it exists, and that it is a very powerful tool for the analysis and design of digital filters \cite{bib:OppenheimDSP}. What should be retained at this stage is that, given the transfer function of a digital filter in the \(Z\)-domain, its frequency response is obtained simply by substituting \(Z\) with \(e^{j2\pi\Omega}\): the frequency response of a digital filter is the Z-transform evaluated on the~unit circle of the complex plane, as illustrated in Fig.~\ref{fig:z_unit_circle}.

As for analogue filters, digital filters also have zeros and poles, as shown in the factorised form of Eq.~(\ref{eq:HZ_general}) above. As in the analogue case, the total response on a Bode diagram is the sum of the~individual contributions of each zero and pole. Unlike for analogue filters, however, the contribution of a~single digital zero or pole does not reduce to simple straight-line asymptotes: the normalised frequency \(\Omega\) is periodic and bounded, unlike the unbounded frequency \(f\) of the analogue case, so there is no far-away regime in which the response can be approximated by a straight line.

\begin{figure}[!ht]
    \centering
    \includegraphics[width=0.6\textwidth]{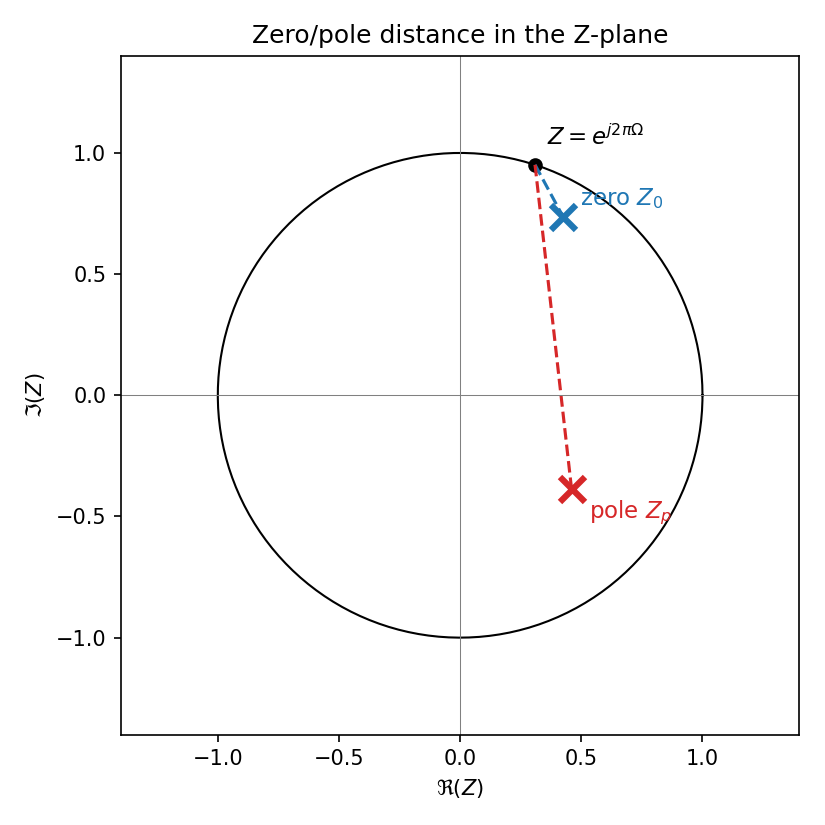}
    \caption{Geometric interpretation of the magnitude response of a single zero and pole in the digital domain: the~contribution at a given frequency \(\Omega\) is given by the distance between \(Z=e^{j2\pi\Omega}\) on the unit circle and the zero (blue) or pole (red).}
    \label{fig:z_plane_distance}
\end{figure}

Substituting back \(Z=e^{j2\pi\Omega}\) in the transfer function, however, shows that the magnitude response contributed by a single zero or pole is simply given by the \emph{distance}, in the complex plane, between the~point \(Z=e^{j2\pi\Omega}\) and the zero or pole itself: the closer a zero lies to the unit circle, the more strongly it attenuates the frequencies close to its angle, and, similarly, the closer a pole lies to the unit circle, the more strongly it amplifies the frequencies close to its angle. This is illustrated in Fig.~\ref{fig:z_plane_distance}, where the~contribution of a zero \(Z_0\) and a pole \(Z_p\) at a given evaluation frequency \(\Omega\) are represented by the~length of the dashed segments joining them to the point \(Z=e^{j2\pi\Omega}\) on the unit circle. For clarity, only one zero and one pole are shown; in practice, since a physical filter has real-valued coefficients, any zero or pole off the real axis cannot appear on its own and must always be paired with its complex conjugate, reflected across the~real axis, with the total response given by the sum of the (equal) contributions of both members of each pair -- as discussed further, for poles, in Section~4.5.

\subsection{FIR filters}

Digital filters are commonly classified into two main categories: \textbf{\textit{finite impulse response (FIR)}} and \textbf{\textit{infinite impulse response (IIR)}} filters. This section focuses on FIR filters, and IIR filters will be covered separately.

FIR filters contain only feedforward terms, i.e.\ they are a linear combination of the current and past input samples. Setting all feedback coefficients \(a_i\) to zero in Eq.~(\ref{eq:general_difference_equation}), the transfer function of an FIR filter can be written as

\begin{equation}
y[n] = \sum_{i=0}^{N-1} b_i\, x[n-i], \qquad H(Z) = \sum_{i=0}^{N-1} b_i\, Z^{-i}.
\label{eq:fir_definition}
\end{equation}

Since there is no feedback term, the impulse response of the filter is simply given by its coefficients, \(h[n]=b_n\), and is therefore exactly zero after the \(N\) samples following the input impulse -- hence the name \emph{finite} impulse response. FIR filters are, as a consequence, always stable: for a bounded input, the output at any given time is a sum of a finite number of bounded terms, and a finite sum of bounded terms is itself always bounded -- it can therefore never diverge, irrespective of the values chosen for the~coefficients \(b_i\).

In the factorised form of Eq.~(\ref{eq:HZ_general}), an FIR filter (with no feedback part, i.e.\ \(M=1\)) reduces to \(H(Z) = b_0\prod_{i=1}^{N-1}\left(1-Z_{0i}Z^{-1}\right) = b_0\,Z^{-(N-1)}\prod_{i=1}^{N-1}\left(Z-Z_{0i}\right)\): an \((N-1)\)-fold pole at the origin, \(Z=0\), together with the \(N-1\) explicit zeros \(Z_{0i}\). Since the origin is equidistant from every point of the~unit circle, this pole contributes only a constant, frequency-independent scaling factor to the magnitude response, and plays no role whatsoever in shaping it: the frequency response of an FIR filter is therefore governed entirely by the location of its zeros, as illustrated in the example of Section~4.2.2 below.

\begin{figure}[!ht]
    \centering
    \includegraphics[width=0.85\textwidth]{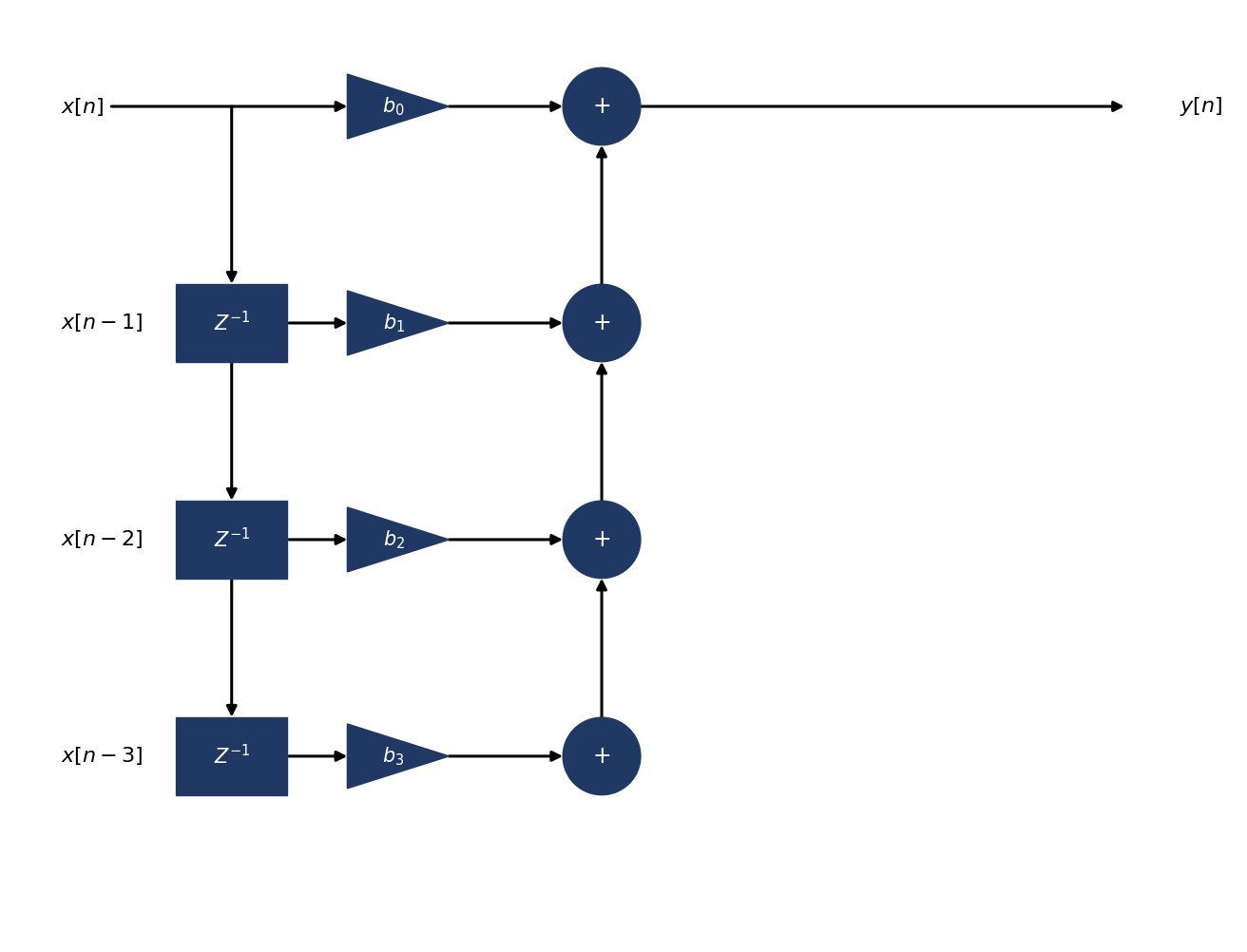}
    \caption{Block diagram of a generic 4-tap FIR filter: the input \(x[n]\) is passed through a chain of unit delays, each tap is scaled by its coefficient \(b_i\), and the scaled taps are summed to produce the output \(y[n]\).}
    \label{fig:fir_block_diagram}
\end{figure}

In hardware, Eq.~(\ref{eq:fir_definition}) is naturally implemented as a tapped delay line, as illustrated in Fig.~\ref{fig:fir_block_diagram}: the~input is passed through a chain of unit delays (\(Z^{-1}\)), a copy of the signal is tapped off after each delay, each tap is scaled by the corresponding coefficient \(b_i\), and the scaled taps are finally summed to form the~output.

FIR filters can be designed to have an exactly linear phase response, meaning that every frequency component of the signal is delayed by the same amount of time, and the filter therefore introduces no phase distortion -- a property obtained simply by choosing coefficients that are symmetric about the centre of the filter, \(b_i=b_{N-1-i}\). This can be seen by pairing each coefficient with its symmetric partner in Eq.~(\ref{eq:fir_definition}):
\begin{equation}
b_i\, Z^{-i} + b_i\, Z^{-(N-1-i)} = b_i\, Z^{-\frac{N-1}{2}}\left(Z^{\frac{N-1}{2}-i} + Z^{-\left(\frac{N-1}{2}-i\right)}\right),
\end{equation}
which, evaluated on the unit circle \(Z=e^{j2\pi\Omega}\), becomes \(2b_i\cos\!\big(2\pi\Omega(\tfrac{N-1}{2}-i)\big)\,e^{-j2\pi\Omega\frac{N-1}{2}}\), i.e.\ a purely real quantity times the common factor \(e^{-j2\pi\Omega\frac{N-1}{2}}\). Every such pair -- and hence the full sum \(H(e^{j2\pi\Omega})\) -- therefore shares this same linear-phase factor, multiplying a purely real function of \(\Omega\): the phase is exactly linear in \(\Omega\), corresponding to a constant group delay of \((N-1)/2\) samples, with no other frequency-dependent phase contribution. An antisymmetric choice of coefficients, \(b_i=-b_{N-1-i}\), gives an entirely analogous result -- each pair again factorises into the same common delay term \(e^{-j2\pi\Omega\frac{N-1}{2}}\), only with the cosine above replaced by a sine -- so the phase remains exactly linear in \(\Omega\), aside from a fixed \(90^\circ\) offset \cite{bib:OppenheimDSP,bib:Lyons}. This is particularly valuable in applications where the shape of the signal matters, such as audio processing or beam instrumentation waveforms. FIR filters can also be designed to approximate essentially arbitrary frequency responses: since, as noted above, an FIR filter's coefficients are simply its impulse response, \(h[n]=b_n\), this is done by starting from the impulse response corresponding to the desired, target frequency response, and using its -- possibly truncated -- samples directly as the tap coefficients \(b_i\). This makes FIR filters very flexible. This flexibility, however, comes at a price: FIR filters typically require significantly more coefficients (and therefore more computational resources) than an equivalent IIR filter to achieve the same level of frequency selectivity, especially for narrow-band responses.

\begin{figure}[!ht]
    \centering
    \includegraphics[width=0.95\textwidth]{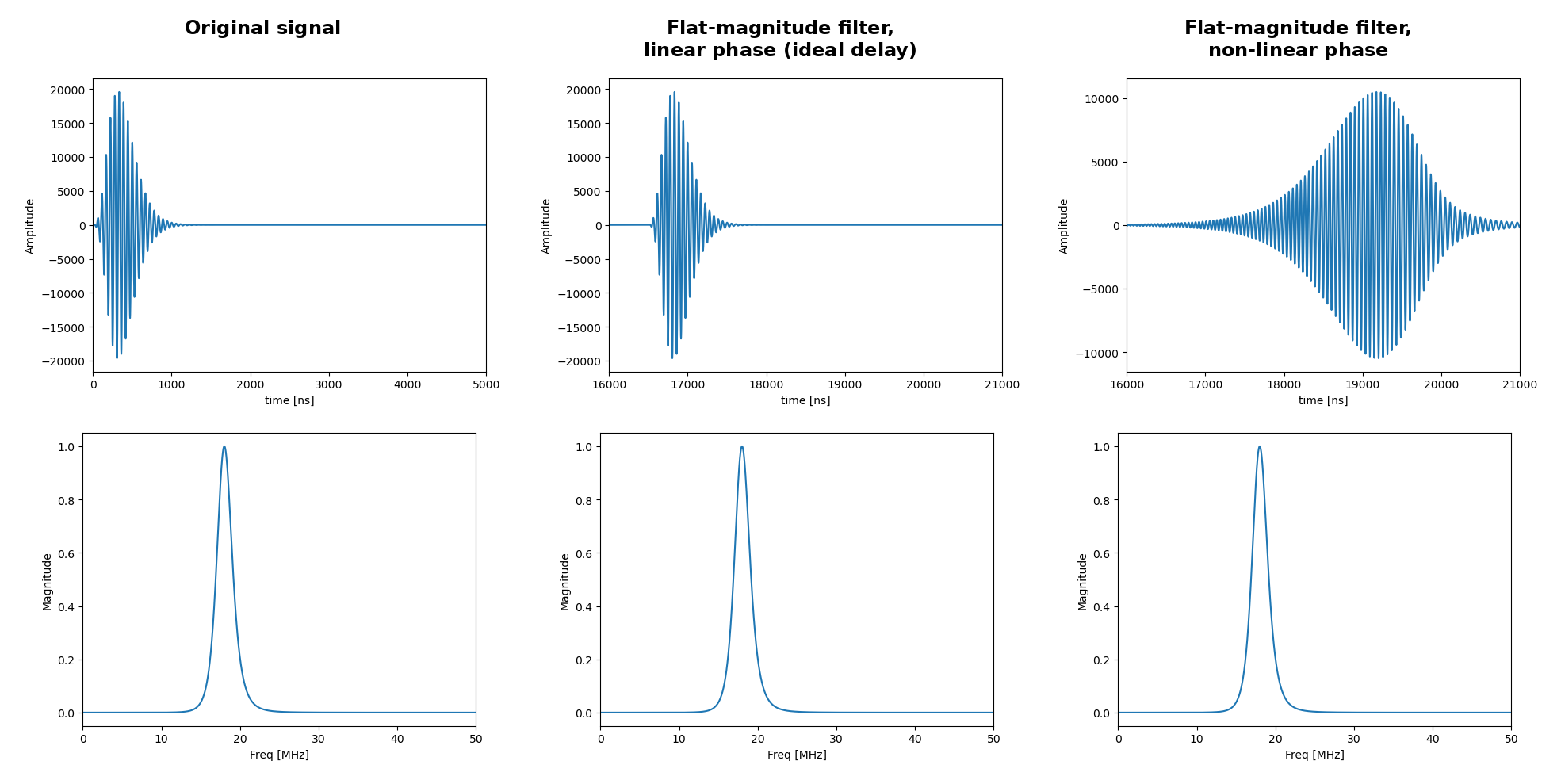}
    \caption{Effect of phase (non-)linearity on the shape of a signal, for two filters with the same flat magnitude response (bottom row) but different phase response. A short resonant burst (left) is passed through a filter with linear phase, which behaves as an ideal delay line and simply delays the burst without changing its shape (centre), and through a filter with non-linear phase, which spreads the burst out over a much longer duration despite leaving its magnitude spectrum unchanged (right).}
    \label{fig:fir_phase_distortion}
\end{figure}

The practical consequence of a non-linear phase response is illustrated in Fig.~\ref{fig:fir_phase_distortion} for a short resonant burst centred around 18\,MHz. Two filters are compared, both with exactly the same, flat magnitude response over the band of interest -- so that neither of them alters the magnitude spectrum of the burst -- but with different phase responses: one has an exactly linear phase, behaving as an ideal delay line, while the other does not. Because its phase is linear, the first filter delays every frequency component of the burst by the same amount, so its output is simply a delayed -- but otherwise undistorted -- copy of the input, exactly as if the signal had been passed through a pure delay line. The second filter, despite having the very same magnitude response, delays different frequency components of the burst by different amounts; its output is therefore not simply delayed but stretched out considerably in time, with the~energy of the original short burst spread over a much longer duration. This shows that the magnitude spectrum alone does not determine how a filter affects a signal: a non-linear phase response can severely distort its shape even when the magnitude response is left completely unchanged.

In summary, FIR filters:
\begin{itemize}
\item have only explicit zeros;
\item have an intrinsically time-limited (finite-length) impulse response, equal to their coefficients;
\item are always stable, for any choice of coefficients;
\item can be designed to have an exactly linear phase response;
\item can approximate an arbitrary frequency response, at the cost of a potentially large number of coefficients, particularly for narrow-band filters.
\end{itemize}

\subsubsection{Cancelling a narrow-band disturbance}

FIR filters can be designed to have an exact notch at a specific frequency, which makes them very useful for cancelling narrow-band disturbances in a signal -- for instance an interference at a known, fixed frequency. To achieve this, the zeros of the filter are placed directly on the unit circle, at the angle corresponding to the normalised frequency of the disturbance: the closer the zeros are to the unit circle, the more effective the filter is at cancelling the disturbance, as shown in Section~4.1.

The \textbf{\textit{tune}} of a circular accelerator is the natural frequency of the transverse (betatron) oscillations of the beam around its ideal orbit, normalised to the revolution frequency -- i.e.\ the number of such oscillations completed per turn -- and is determined by the focusing optics of the machine \cite{bib:Wolski}. A whole number of oscillations per turn is indistinguishable from a fixed observation point, so what is actually measured at a single point of the accelerator, such as a beam position monitor, is only the \textbf{\textit{fractional tune}} \(q\), the fractional part of the tune.

As a purely hypothetical illustration -- not a real, documented case -- imagine that a tune measurement in the CERN LHC is affected by a strong disturbance at 20~kHz originating from nearby power converters -- for instance because the tune-system signal cables happen to be routed too close to the~power-converter cables. The tune system samples at the LHC revolution frequency, \(f_s=11.3\)~kHz; being higher than the sampling frequency, the 20~kHz disturbance is aliased into the base-band (see Section~2.2.1), reappearing as a tone at

\begin{equation}
f_{\mathrm{alias}} = 2f_s - f_{\mathrm{dist}} = 2\times 11.3 - 20 = 2.6~\mathrm{kHz}.
\end{equation}

Fortunately, the fractional tune in this example based on LHC parameters is close to \(q=0.31\), so that the disturbance, at a normalised frequency of \(2.6/11.3 \approx 0.23\), remains reasonably well separated from the tune line. Nevertheless, such a peak can interfere with tune-detection algorithms, and it is preferable to remove it. While the obvious long-term solution is to re-route the offending cables, in the meantime a simple notch, implemented as a second-order FIR filter, can be used. Since a real-valued (physical) filter requires complex-conjugate zeros, the two zeros are placed at

\begin{equation}
Z_{0,1} = e^{j2\pi \times 0.23}, \qquad Z_{0,2} = e^{-j2\pi \times 0.23},
\end{equation}

giving the transfer function

\begin{equation}
H(Z) = Z^{-2}(Z-Z_{0,1})(Z-Z_{0,2}) = 1 - 2\cos(2\pi\times 0.23)\, Z^{-1} + Z^{-2} \approx 1 - \tfrac{1}{4}Z^{-1} + Z^{-2},
\end{equation}

corresponding to the very simple, 3-tap notch filter

\begin{equation}
y[n] = x[n] - \tfrac{1}{4}x[n-1] + x[n-2].
\end{equation}

\begin{figure}[!ht]
    \centering
    \includegraphics[width=1.0\textwidth]{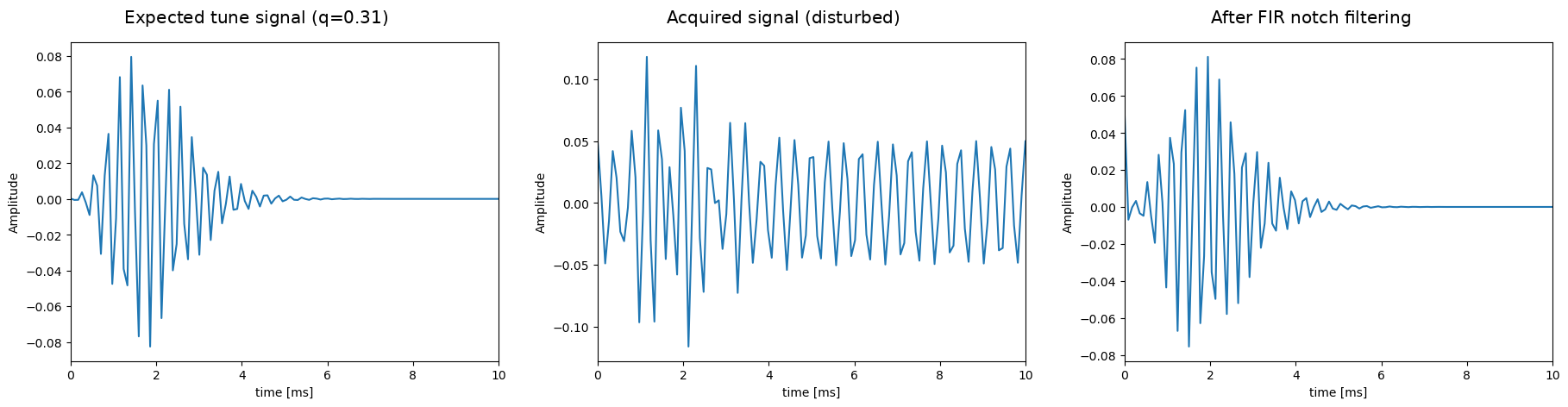}
    \caption{Application of the 3-tap notch filter to an LHC tune measurement: expected tune signal (left), acquired signal contaminated by the 2.6~kHz aliased disturbance (centre), and signal recovered after filtering (right).}
    \label{fig:tune_disturbance}
\end{figure}

Figure~\ref{fig:tune_disturbance} shows the expected, clean tune signal, the actual acquired signal contaminated by the~2.6~kHz disturbance, and the result of applying this simple notch filter, which is enough to recover a~signal very close to the expected one.

\subsubsection{High-frequency noise removal}

\begin{figure}[!ht]
    \centering
    \includegraphics[width=1.0\textwidth]{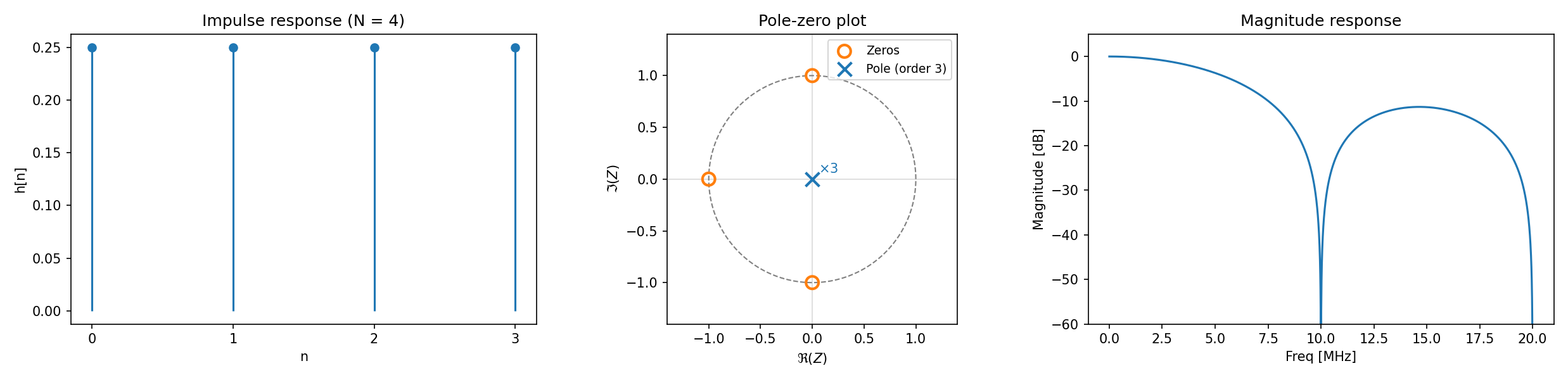}
    \caption{Impulse response (left), pole-zero plot (centre), and magnitude response (right, linear frequency axis) of the 4-tap moving-average FIR filter, sampled at \(f_s = 40\)~MHz. The magnitude response exhibits the expected low-pass behaviour, with nulls at multiples of \(f_s/4 = 10\)~MHz, directly beneath the three zeros shown in the centre panel.}
    \label{fig:fir_movingaverage}
\end{figure}

One of the most common filters is the simple average. Averaging is known to remove noise -- or, more precisely, high-frequency noise -- while preserving the slowly varying part of a signal. The equation for an average over four samples is

\begin{equation}
y[n] = \frac{1}{4}\Big(x[n] + x[n-1] + x[n-2] + x[n-3]\Big),
\end{equation}

which is immediately recognisable as an FIR filter with \(N=4\) taps, and, in the factorised form of Eq.~(\ref{eq:HZ_general}), can be written as \(H(Z) = \tfrac{1}{4}\,Z^{-3}(Z-j)(Z+1)(Z+j)\): the triple pole at the origin common to every 4-tap FIR filter (see above), together with three explicit zeros, at \(Z=j\), \(Z=-1\) and \(Z=-j\) -- i.e.\ on the unit circle, at the normalised frequencies \(\Omega=1/4\), \(1/2\) and \(3/4\) respectively. This pole-zero configuration is shown in the centre panel of Fig.~\ref{fig:fir_movingaverage}. Recalling the distance interpretation of Section~4.1, each of the three zeros only attenuates the frequencies in its immediate neighbourhood, and the response would rise again away from it; the pole at the origin, as noted above, plays no role in shaping the response. By spacing the three zeros evenly around the circle, the attenuated neighbourhood of one zero adjoins that of the next, so that the response stays low at essentially every frequency other than the~vicinity of \(\Omega=0\), where no zero is present. The result is the low-pass magnitude response shown in the~right panel of Fig.~\ref{fig:fir_movingaverage}, together with the corresponding impulse response (left panel), simply given by the~four constant coefficients of the filter.

Unlike the Bode-style plots used earlier for the individual analogue pole and zero contributions (Fig.~\ref{fig:bode_zero_pole_family}), which use a logarithmic frequency axis well suited to displaying asymptotic straight-line behaviour over many decades, the magnitude response in the right panel of Fig.~\ref{fig:fir_movingaverage} -- as in most digital magnitude response plots in this text -- uses a linear frequency axis: the normalised frequency \(\Omega\) is bounded and, unlike the unbounded analogue frequency \(f\), includes the point \(\Omega=0\), which cannot be represented on a logarithmic scale. The magnitude itself, however, continues to be expressed in dB, exactly as in the analogue case.

\subsection{IIR filters}

Infinite impulse response (IIR) filters are digital filters whose difference equation, Eq.~(\ref{eq:general_difference_equation}), also includes the feedback part, i.e.\ at least one non-zero coefficient \(a_i\). It is precisely because of this feedback -- the~contribution of past outputs to the current one -- that the impulse response of these filters can, in general, extend to infinity, hence their name.

Unlike FIR filters, whose poles are all located at the origin (as can be seen from the leading \(Z^{-(N-1)}\) factor in their transfer function), IIR filters have explicit, non-trivial poles, and, just like their analogue counterparts, can therefore easily implement resonant and sharply selective low-pass behaviour.

\begin{figure}[!ht]
    \centering
    \includegraphics[width=1.0\textwidth]{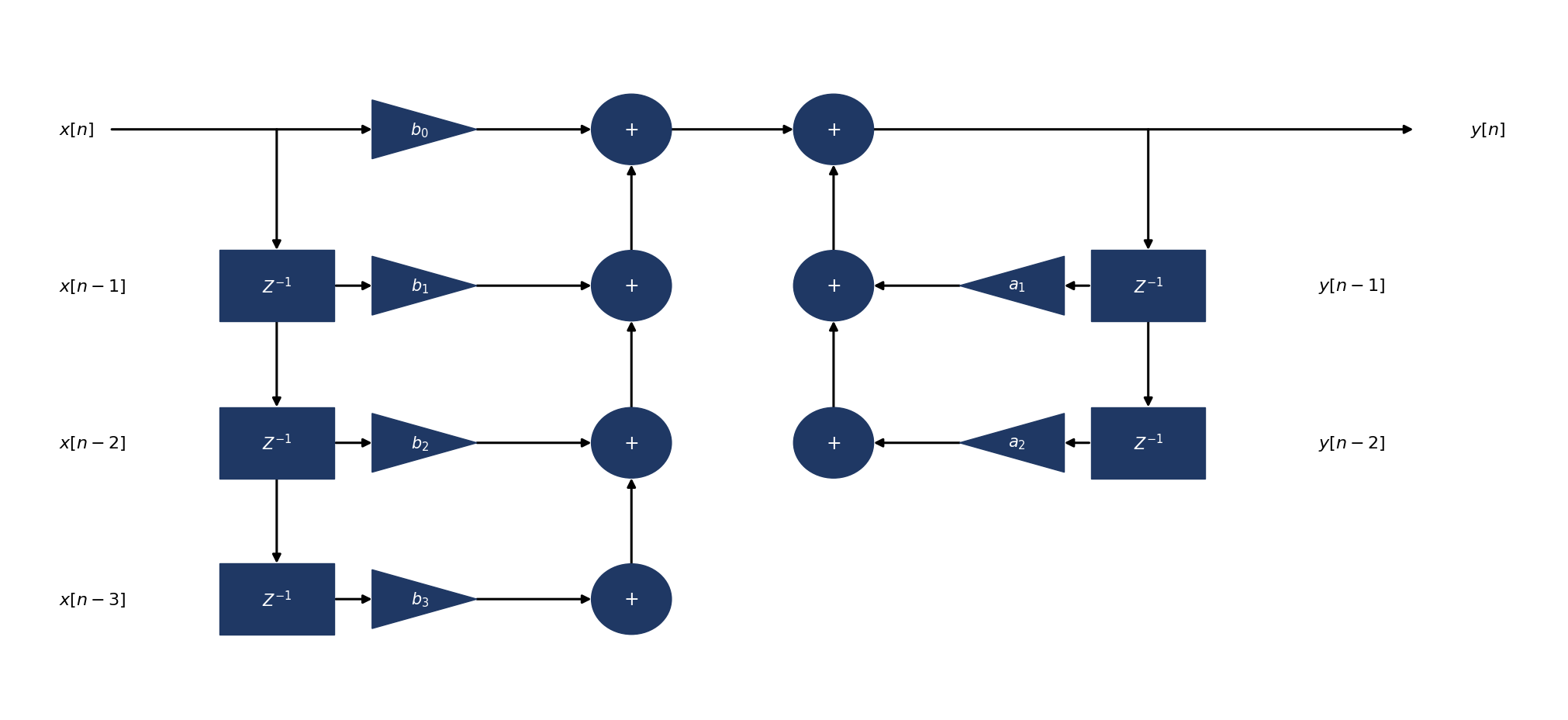}
    \caption{Direct Form I realisation of a general IIR filter (here with \(N=4\) feedforward taps and \(M-1=2\) feedback taps): the feedforward part (left) is identical to an FIR filter, its output is combined with a separate feedback part (right), built from past output samples, to produce \(y[n]\).}
    \label{fig:direct_form_I}
\end{figure}

An IIR filter can be implemented in more than one equivalent structure. The most direct one, known as \textbf{\textit{Direct Form I}}, simply implements Eq.~(\ref{eq:general_difference_equation}) as written: a feedforward tapped delay line, identical to the one used for FIR filters, feeding an adder, combined with a separate feedback tapped delay line built from the past output samples, as shown in Fig.~\ref{fig:direct_form_I}. Each branch, associated with one coefficient \(a_i\) or \(b_i\) and one delayed sample, is referred to as a \textbf{\textit{tap}}; Direct Form I therefore requires \(N-1\) feedforward and \(M-1\) feedback delay elements.

\begin{figure}[!ht]
    \centering
    \includegraphics[width=0.85\textwidth]{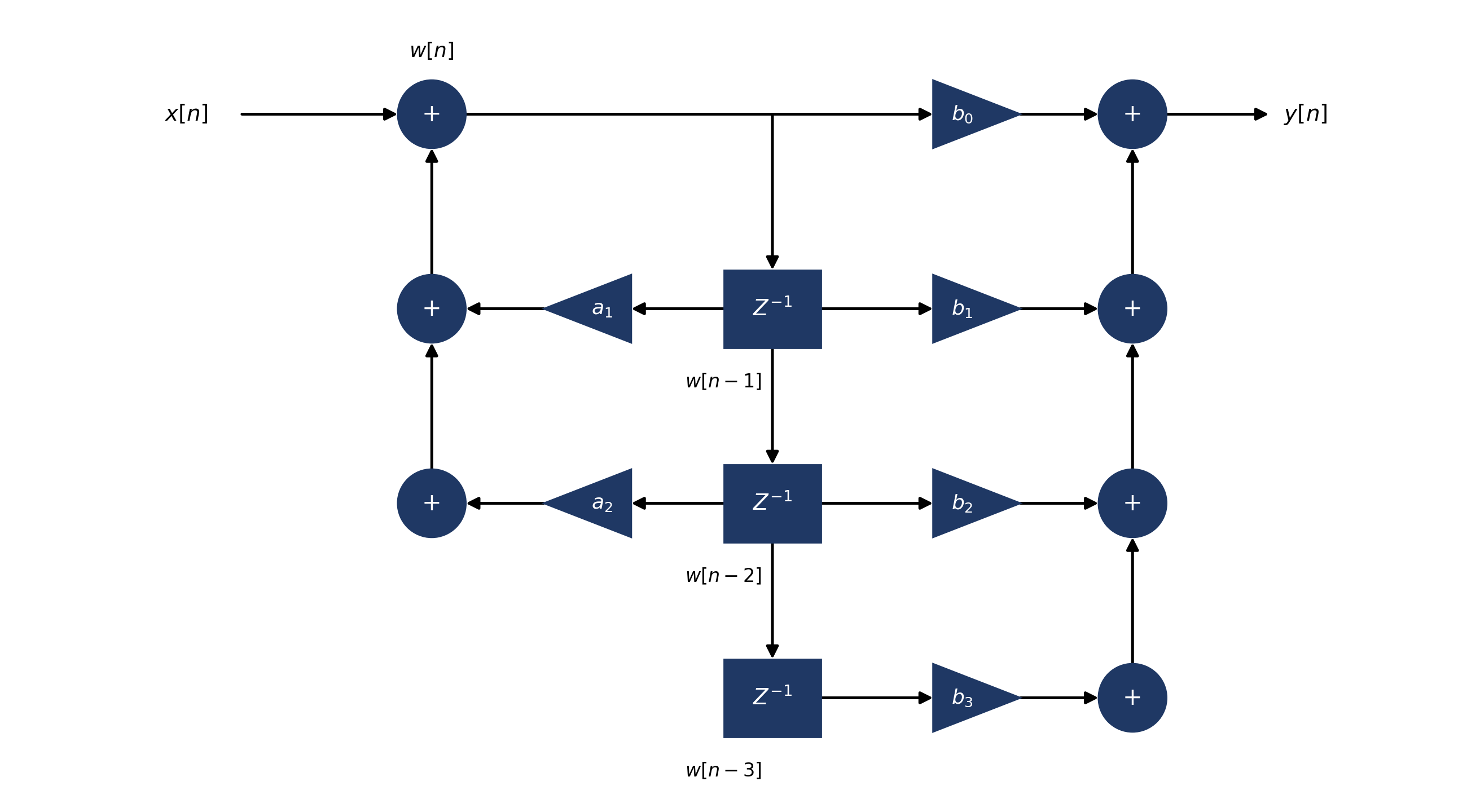}
    \caption{Direct Form II (canonical) realisation of the same IIR filter: a single shared delay line, built from the~intermediate sequence \(w[n]\), feeds both the feedback taps \(a_i\) (left, producing \(w[n]\)) and the feedforward taps \(b_i\) (right, producing \(y[n]\)), roughly halving the number of delay elements needed compared to Direct Form I.}
    \label{fig:direct_form_II}
\end{figure}

A more economical structure, known as \textbf{\textit{Direct Form II}} and shown in Fig.~\ref{fig:direct_form_II}, follows from a~simple observation: Direct Form I applies the feedforward part and the feedback part one after the~other, and, since both are LTI systems, their order can be swapped without changing the overall filter (Section~3.3). Applying the feedback part first, directly to \(x[n]\), gives an intermediate sequence \(w[n] = x[n] + \sum_{i=1}^{M-1} a_i\, w[n-i]\); the feedforward part is then applied to \(w[n]\) instead of \(x[n]\), giving \(y[n] = \sum_{i=0}^{N-1} b_i\, w[n-i]\). The delayed samples \(w[n-i]\) it needs are exactly the ones already stored by the feedback part's own delay line, so a single shared delay line, of length \(\max(N,M)-1\), suffices for both parts -- roughly halving the number of delay elements needed compared to Direct Form I.

The main advantage of IIR filters is their ability to implement very selective low-pass and resonant responses with a much smaller number of taps than an equivalent FIR filter, as illustrated in the example below. IIR filters, however, also have drawbacks: they cannot, in general, be implemented with an exactly linear phase response, which can be an issue, or at least something to keep in mind, in applications sensitive to signal distortion; and, because of the feedback, unlike FIR filters they are not guaranteed to be stable. This last point is addressed later in this section.

\subsection{IIR example: average approximation}

Let us consider an example taken from beam position monitoring in the LHC. The LHC orbit system has a bandwidth of 50~Hz, while the revolution frequency is 11.3~kHz. Suppose we are asked to implement the orbit filter -- the filter that turns the bunch-by-bunch, turn-by-turn samples available from the BPM system into a 50~Hz data stream -- as a simple moving average, so that it can be sampled at any time or used to immediately detect variations. Even assuming the machine is filled with a single bunch, this would require on the order of 225 taps; for a fully filled machine, this number would grow to more than 500k taps. This would clearly consume a prohibitive amount of resources, so let us see whether this can be reduced.

The key observation is that an \(N\)-sample boxcar sum,

\begin{equation}
y[n] = \sum_{i=0}^{N-1} x[n-i],
\end{equation}

can be rewritten recursively as

\begin{equation}
y[n] = x[n] + y[n-1] - x[n-N].
\label{eq:boxcar_recursive}
\end{equation}

The new running sum is simply the previous one, plus the newest sample, minus the sample leaving the window. This already reduces the required arithmetic to a single addition and subtraction per sample, although a delay line of length \(N\) is still needed to retain \(x[n-N]\). Since \(x[n-N]\) is itself just the average we are trying to extract, plus noise, it can be approximated by the last computed average, \(x[n-N] \approx y[n-1]/N\); substituting this into Eq.~(\ref{eq:boxcar_recursive}) removes the need for this delay line entirely, turning the filter into the first-order recursive (leaky) average

\begin{equation}
\tilde{y}[n] = x[n] + \frac{N-1}{N}\, \tilde{y}[n-1].
\label{eq:leaky_average}
\end{equation}

This filter now requires only a single tap, and the number of coefficients and delay elements no longer depends on \(N\) at all.

\begin{figure}[!ht]
    \centering
    \includegraphics[width=1.0\textwidth]{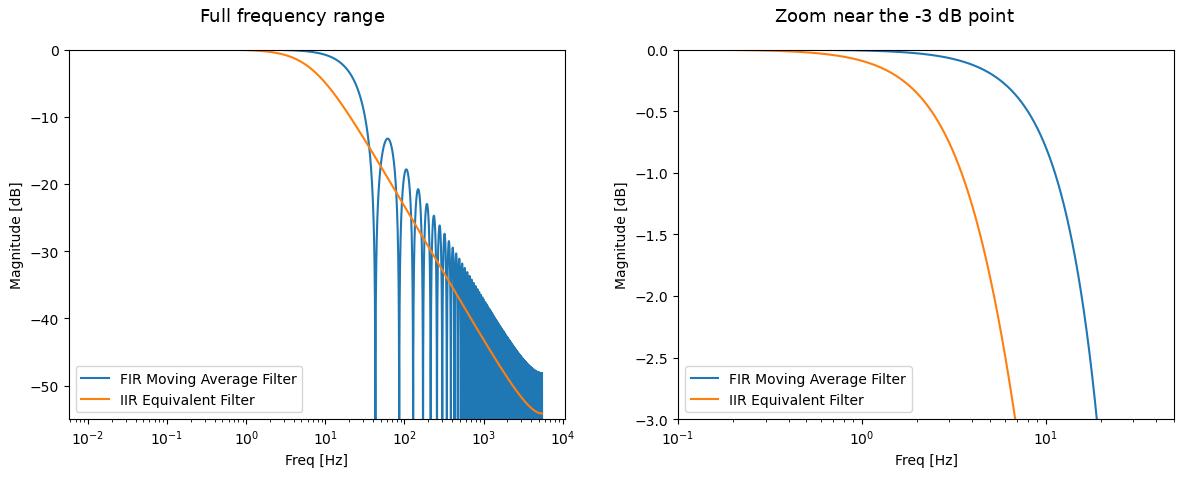}
    \caption{Magnitude response of the exact, 225-tap FIR moving-average filter compared to its single-tap IIR approximation: full frequency range (left), and zoom near the \(-3\)~dB point (right), showing the small discrepancy between the two cut-off frequencies.}
    \label{fig:iir_vs_fir_average}
\end{figure}

Comparing the frequency response of this IIR approximation to that of the exact FIR moving average it replaces (Fig.~\ref{fig:iir_vs_fir_average}, here shown with a logarithmic frequency axis, unlike Fig.~\ref{fig:fir_movingaverage} above, since the much longer, 225-tap filter has many more nulls, spread over a much wider frequency range that is more clearly displayed on a log scale), two differences stand out. First, the \(-3\)~dB points of the two filters do not coincide exactly; this is a minor discrepancy, easily corrected by slightly adjusting the feedback coefficient. Second, and more fundamentally, the FIR filter completely nulls out many frequencies outside the pass-band (the region of the spectrum the filter is designed to preserve), while the IIR approximation only attenuates them smoothly, without ever reaching an exact null.

In some cases, such exact notches are actually desirable -- for instance to remove the 50~Hz ripple picked up from power supplies -- and can be reintroduced by adding back a pair of zeros with a notch behaviour: with the sampling frequency here equal to the LHC revolution frequency, \(f_s=11.3\)~kHz, the~50~Hz ripple corresponds to the normalised frequency \(\Omega_{50}=50/11300\approx4.4\times10^{-3}\), placing the notch zeros at \(Z_{0,1,2}=e^{\pm j2\pi\Omega_{50}}\).

Simply adding this pair of zeros, however, is not enough here: the orbit filter's pass-band extends, by design, right up to 50~Hz, so the notch frequency \(\Omega_{50}\) coincides almost exactly with the filter's own \(-3\)~dB point. A bare pair of zeros, with no pole nearby to confine its influence, does not settle back to a~flat response away from its own frequency -- its magnitude keeps growing with distance from the zeros, all the way to the opposite side of the unit circle, exactly like the rising zero contribution of Fig.~\ref{fig:bode_zero_pole_family}. Placed directly at the edge of the pass-band, such a zero pair would not just remove a thin sliver of spectrum at 50~Hz: it would erode the pass-band itself and let the response climb well above unity beyond it, defeating the purpose of the filter.

To keep the notch narrow, a complex-conjugate pair of poles is added at the same angle as the~zeros, but pulled slightly inside the unit circle, at \(Z_{p,1,2}=r\,e^{\pm j2\pi\Omega_{50}}\) with \(r\lesssim1\). Since each pole nearly coincides with a zero, their contributions cancel almost everywhere; only in the immediate neighbourhood of \(\Omega_{50}\), where the zero -- sitting exactly on the unit circle -- still forces the response to zero, does the notch remain visible. The closer \(r\) is to unity, the narrower the notch (and, as for the narrow-band filter of Section~5.3, the more sensitive it becomes to the exact value of \(\Omega_{50}\) and to coefficient precision). Writing \(c=2\cos(2\pi\Omega_{50})\), this gives the notch transfer function

\begin{equation}
H_{\mathrm{notch}}(Z) = \frac{1-c\,Z^{-1}+Z^{-2}}{1-rc\,Z^{-1}+r^2 Z^{-2}},
\label{eq:sharpened_notch}
\end{equation}

which, cascaded with the leaky average of Eq.~(\ref{eq:leaky_average}), yields the combined difference equation

\begin{equation}
y[n] = x[n] - c\,x[n-1] + x[n-2] + (a+rc)\,y[n-1] - (r^2+arc)\,y[n-2] + a r^2\, y[n-3],
\label{eq:leaky_average_notched}
\end{equation}

with \(a=(N-1)/N\). Compared to the plain leaky average, this requires two additional feedforward taps (from the notch zeros) and two additional feedback taps (from the compensating poles) -- four additional taps in total, rather than the two that adding the zeros alone would suggest.

\begin{figure}[!ht]
    \centering
    \includegraphics[width=1.0\textwidth]{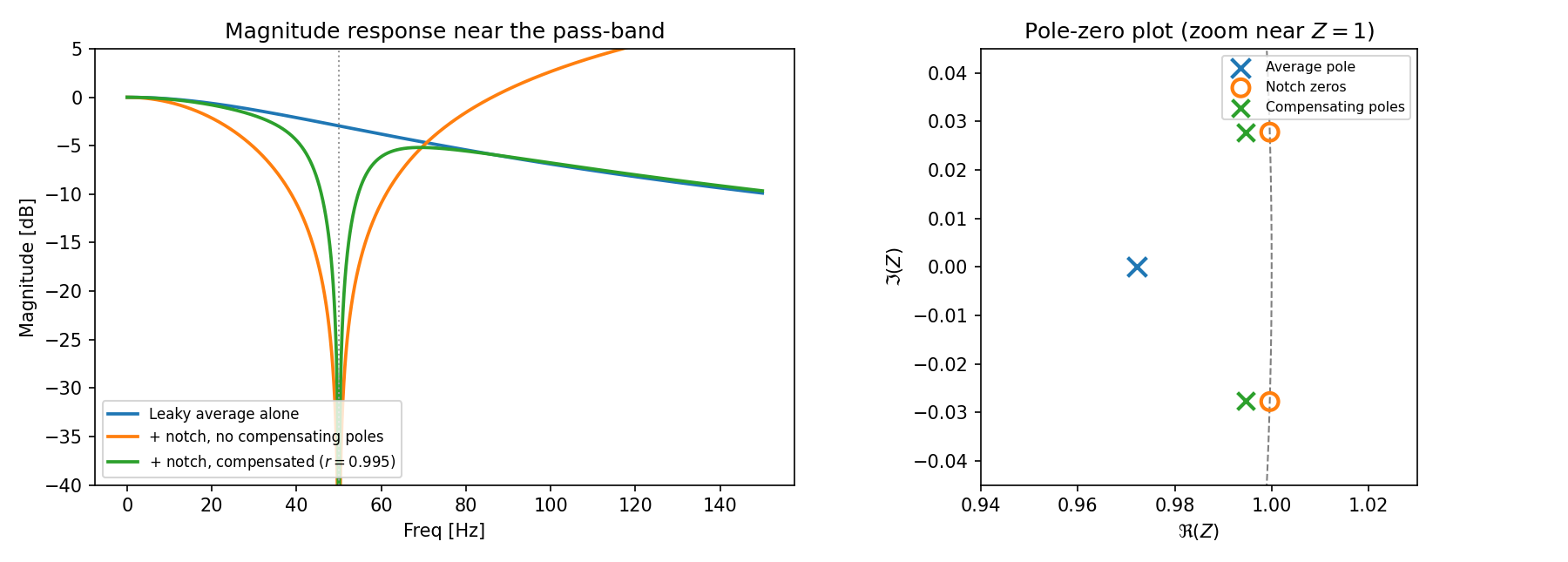}
    \caption{Effect of adding a 50~Hz notch to the leaky-average orbit filter. Left: magnitude response of the plain leaky average, the bare notch (no compensating poles), and the compensated notch (\(r=0.995\)), showing how the~uncompensated zero pair erodes the pass-band and keeps rising beyond it, while the compensated version leaves the pass-band essentially unchanged. Right: pole-zero plot of the compensated filter, zoomed in near \(Z=1\).}
    \label{fig:notched_leaky_average}
\end{figure}

Figure~\ref{fig:notched_leaky_average} compares the magnitude response of the plain leaky average, the bare notch (\(r=0\), i.e.\ no compensating poles), and the compensated notch (\(r=0.995\)). The bare notch visibly erodes the~pass-band well before 50~Hz and keeps rising beyond it, while the compensated notch tracks the plain leaky average almost exactly away from 50~Hz, down to a narrow, deep null exactly at the disturbance frequency. The right-hand panel shows the corresponding pole-zero plot, zoomed in near \(Z=1\): the pass-band pole and the closely-spaced notch zero-pole pair are all clustered at very small angles, reflecting how close the disturbance frequency lies to both the origin and the edge of the pass-band.

\subsection{IIR, unit circle and stability}

The stability condition already anticipated above -- that all poles of \(H(Z)\) must lie strictly inside the unit circle -- is best appreciated through the simplest possible IIR filter, a single-pole recursive filter,

\begin{equation}
y[n] = x[n] + Z_p\, y[n-1].
\end{equation}

\begin{figure}[!ht]
    \centering
    \includegraphics[width=0.75\textwidth]{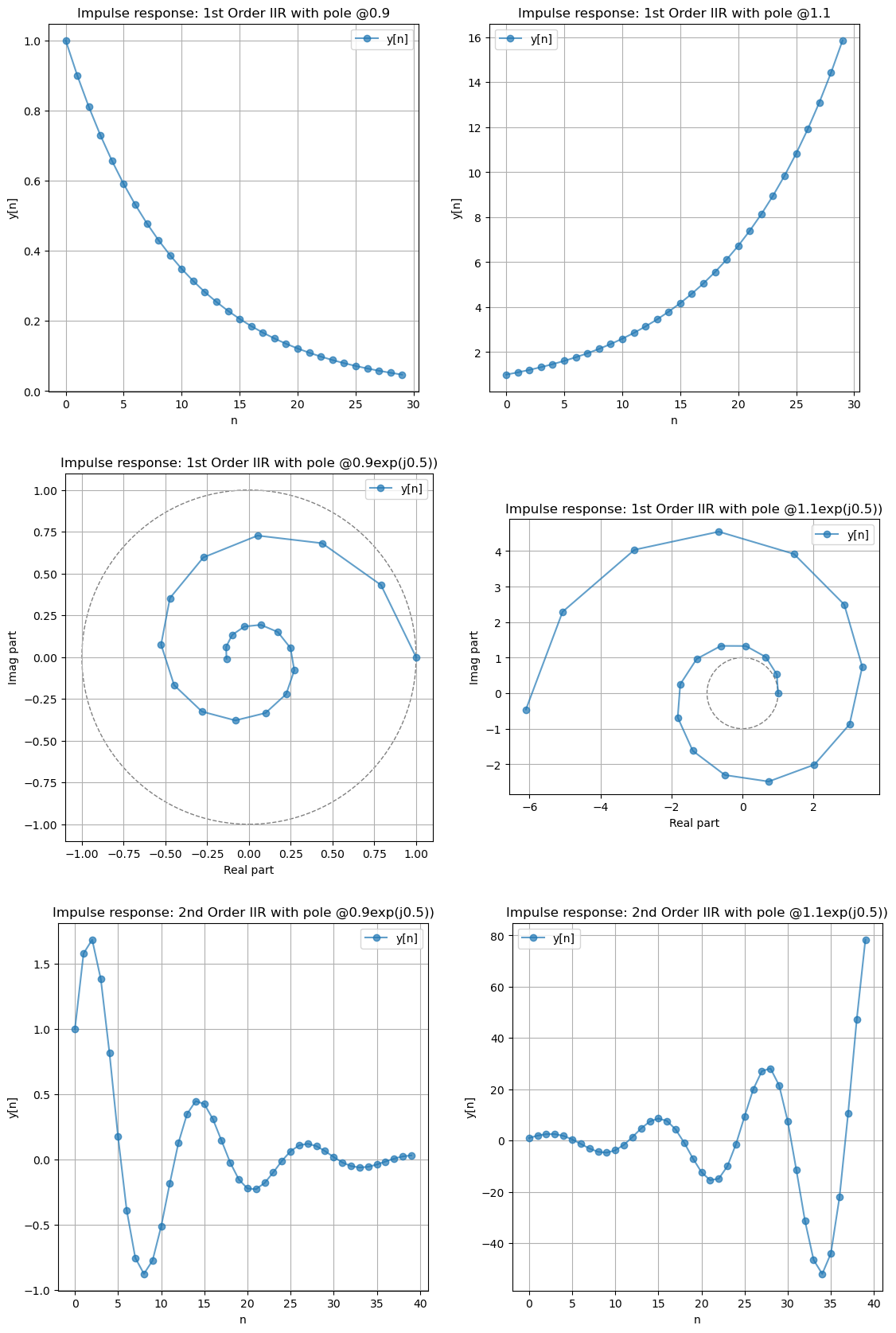}
    \caption{Impulse response of a single-pole IIR filter, \(y[n]=(Z_p)^n\), for a stable (left column) and unstable (right column) pole location. Top: real pole, \(Z_p=0.9\) and \(Z_p=1.1\). Centre: complex pole, \(Z_p=0.9\,e^{j0.5}\) and \(Z_p=1.1\,e^{j0.5}\), shown as a trajectory in the complex plane. Bottom: the corresponding real-valued, second-order filter obtained from the complex-conjugate pole pair, \(y[n]=x[n]-2\,\mathrm{Re}(Z_p)\,y[n-1]+|Z_p|^2 y[n-2]\).}
    \label{fig:iir_pole_gallery}
\end{figure}

Exciting this filter with a unit impulse, \(x[n]=\delta[n]\), gives the impulse response \(y[n] = (Z_p)^n\), whose behaviour depends entirely on the location of the pole \(Z_p\) in the complex plane, as shown in Fig.~\ref{fig:iir_pole_gallery}. For a real pole \(Z_p=0.9\) (inside the unit circle), the impulse response is a simple decaying exponential: the filter is stable. For a real pole \(Z_p=1.1\) (outside the unit circle), the same expression instead grows without bound: the filter is unstable. If the pole is complex, e.g.\ \(Z_p = 0.9\,e^{j0.5}\), the~sequence \(y[n]\) traces a spiral in the complex plane that converges to the origin, corresponding to a damped oscillation and a stable filter; for \(Z_p=1.1\,e^{j0.5}\), the same spiral instead diverges outward, and the filter is unstable.

Since a physical filter must have real-valued coefficients, a complex pole \(Z_p\) must always appear together with its complex conjugate \(Z_p^*\). Combining the pair \(Z_p\), \(Z_p^*\) gives the real, second-order recursive filter

\begin{equation}
y[n] = x[n] - 2\,\mathrm{Re}(Z_p)\, y[n-1] + |Z_p|^2\, y[n-2],
\end{equation}

whose impulse response is, by linearity, simply the sum of the individual responses of the two conjugate poles, \(y[n] = (Z_p)^n + (Z_p^*)^n = 2\,\mathrm{Re}\big((Z_p)^n\big)\): although each term on its own is complex, the imaginary parts of the two conjugate contributions cancel exactly, leaving, as expected for a filter with real coefficients, a purely real output. This impulse response, shown in the bottom row of Fig.~\ref{fig:iir_pole_gallery}, is a damped oscillation for \(|Z_p|=0.9<1\) (stable) and a growing oscillation for \(|Z_p|=1.1>1\) (unstable). This generalises directly to the case of an arbitrary number of poles and zeros, \(H(Z) = K\, Z^{N-M}\prod_i (Z-Z_{0i}) / \prod_i (Z-Z_{pi})\) of Eq.~(\ref{eq:HZ_general}): the filter is stable if, and only if, every pole satisfies \(|Z_{pi}| < 1\).

\subsection{IIR design with the bilinear transformation}

As already mentioned, it is, except for the simple examples given above, considerably harder to design a~digital filter with a desired frequency response directly and intuitively than it is in the analogue domain. This is why it is common practice to instead start from an analogue design, for which a much larger body of literature and ready-made ``cookbook'' solutions is available, and convert it to the digital domain afterwards. Several approaches exist to perform this conversion, the most common of which is the \textbf{\textit{bilinear transformation}}: writing the analogue prototype as a function of the Laplace variable \(s=j2\pi f\), the~exact mapping \(Z=e^{sT}\) between the analogue and digital domains is approximated by

\begin{equation}
s \approx \frac{2}{T}\, \frac{Z-1}{Z+1}.
\end{equation}

A digital IIR filter is then obtained simply by substituting this expression for \(s\) into the analogue transfer function, \(H_a(s) \rightarrow H(Z)\), directly reusing the poles and zeros of a well-established analogue design (Butterworth, Chebyshev, elliptic, \ldots) without having to design the digital filter from scratch \cite{bib:OppenheimDSP,bib:Lyons}.

\section{Examples of common algorithms and processing techniques in beam instrumentation}

This section presents a few examples of common algorithms and processing techniques used in beam instrumentation, illustrating how the concepts introduced in the previous sections are combined in practice. The examples are based on real systems and measurements performed at CERN, simplified to remove unnecessary complexity and to focus on the DSP techniques they are meant to illustrate.

\subsection{Digital upsampling}

The SPS beam position measurement system is nicknamed \textbf{\textit{ALPS}}, standing for \emph{a logarithmic position system}. The very large dynamic range of the beam position signals, of the order of 90~dB, must be covered with a resolution of the order of 0.01~dB, a consequence of the low sensitivity of the electrodes (about 0.1~dB/mm per electrode). To handle this combination of large dynamic range and fine resolution, the~system relies on logarithmic amplifiers to compress the signal before digitisation.

\begin{figure}[!ht]
    \centering
    \includegraphics[width=0.9\textwidth]{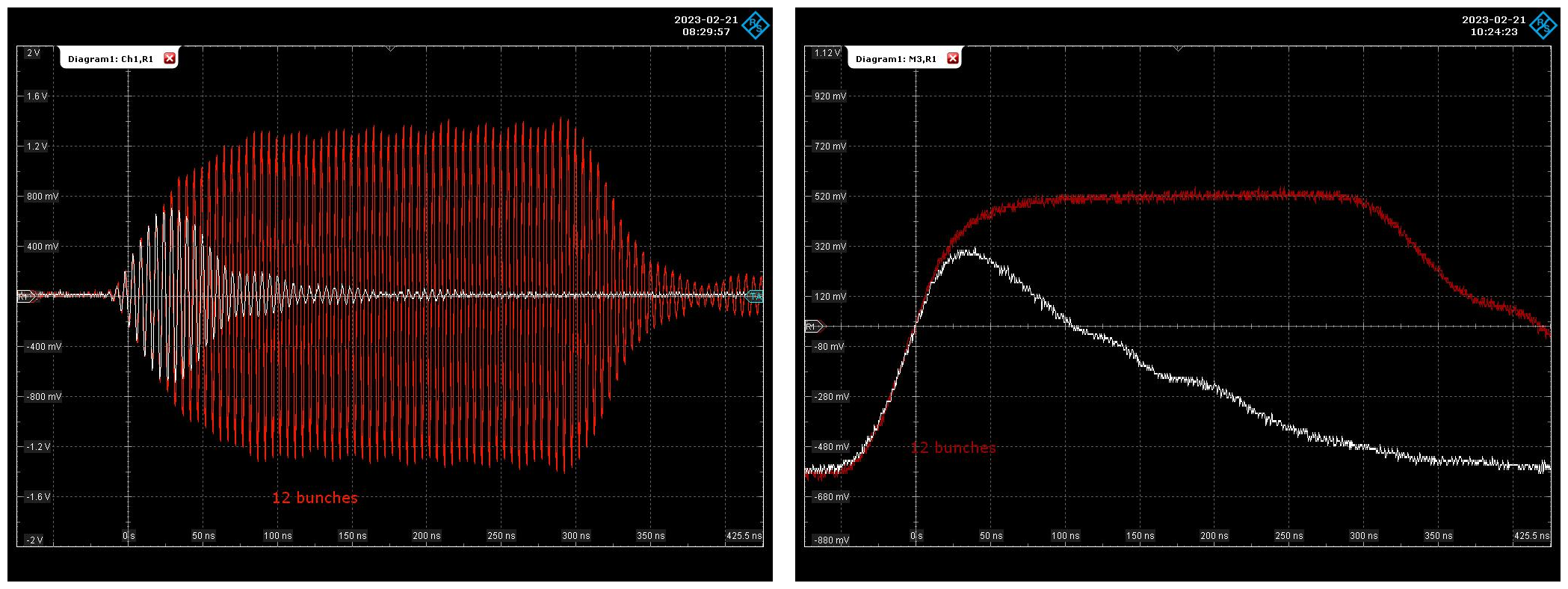}
    \caption{Signal at two points along the ALPS front-end chain, for a single bunch (white) and a train of 12 bunches spaced by 25~ns (red): after the 200~MHz resonant band-pass filter (left) and after the logarithmic amplifier (right).}
    \label{fig:sps_signal_chain}
\end{figure}

The signal picked up by each electrode consists of a short pulse, about 2~ns long, for every bunch; consecutive bunches can be spaced by as little as 5~ns, up to 150~ns \cite{bib:CAS94-LeDuff}. This signal is passed through a~resonant band-pass filter centred at 200~MHz with a bandwidth of about 20~MHz, then through a~logarithmic amplifier, whose output is the envelope of the filtered signal, logarithmically compressed, and finally digitised. Figure~\ref{fig:sps_signal_chain} shows the signal at two points along this front-end chain, for both a single isolated bunch (white) and a train of 12 bunches spaced by 25~ns (red): after the resonant filter (left) and after the logarithmic amplifier (right). Digitisation is performed directly in the tunnel, using a radiation-tolerant ADC running at 40~Msps, with a clock that is free-running, i.e.\ asynchronous with respect to the~beam. The digitised samples are transmitted over optical fibres to the surface, where a~back-end system, based on a field-programmable gate array (FPGA), extracts the pulse amplitude, applies the~electrode calibration, and computes the amplitude difference between electrodes, in dB, from which the~beam position is derived.

For a train of bunches, the logarithmic-amplifier output has a flat top spanning several samples, which is therefore easy to locate and measure reliably. For a single, isolated bunch, however, the pulse is narrow and only a single sample typically falls close enough to its peak; extracting an accurate amplitude in this case requires some additional processing, addressed here using a technique known as \textbf{\textit{digital upsampling}}.

\begin{figure}[!ht]
    \centering
    \includegraphics[width=0.85\textwidth]{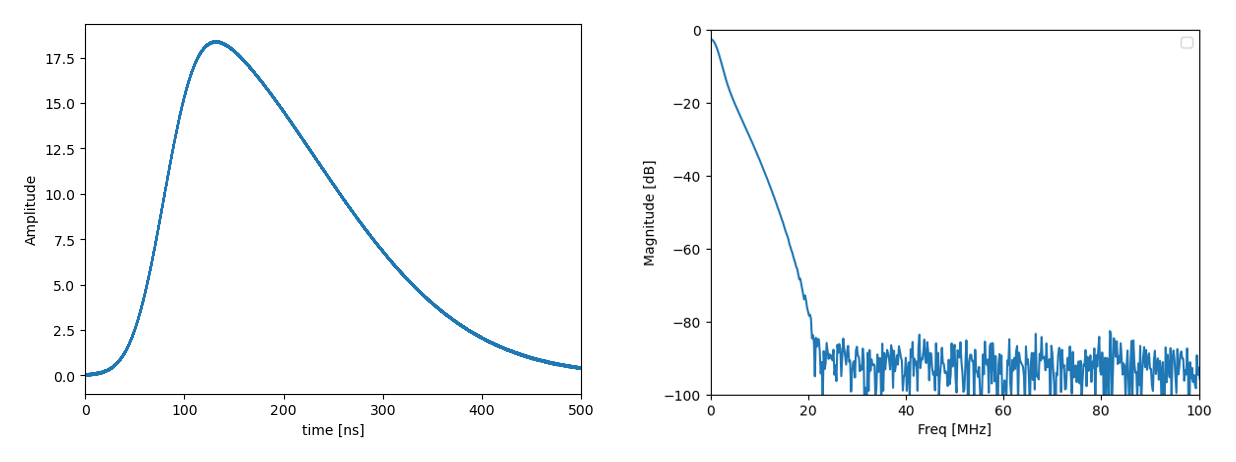}
    \caption{Noiseless, synthesised model of the single-bunch response at the output of the logarithmic amplifier (left) and its magnitude spectrum (right), confirming the expected 20~MHz signal bandwidth.}
    \label{fig:sps_single_bunch_model}
\end{figure}

\begin{figure}[!ht]
    \centering
    \includegraphics[width=0.55\textwidth]{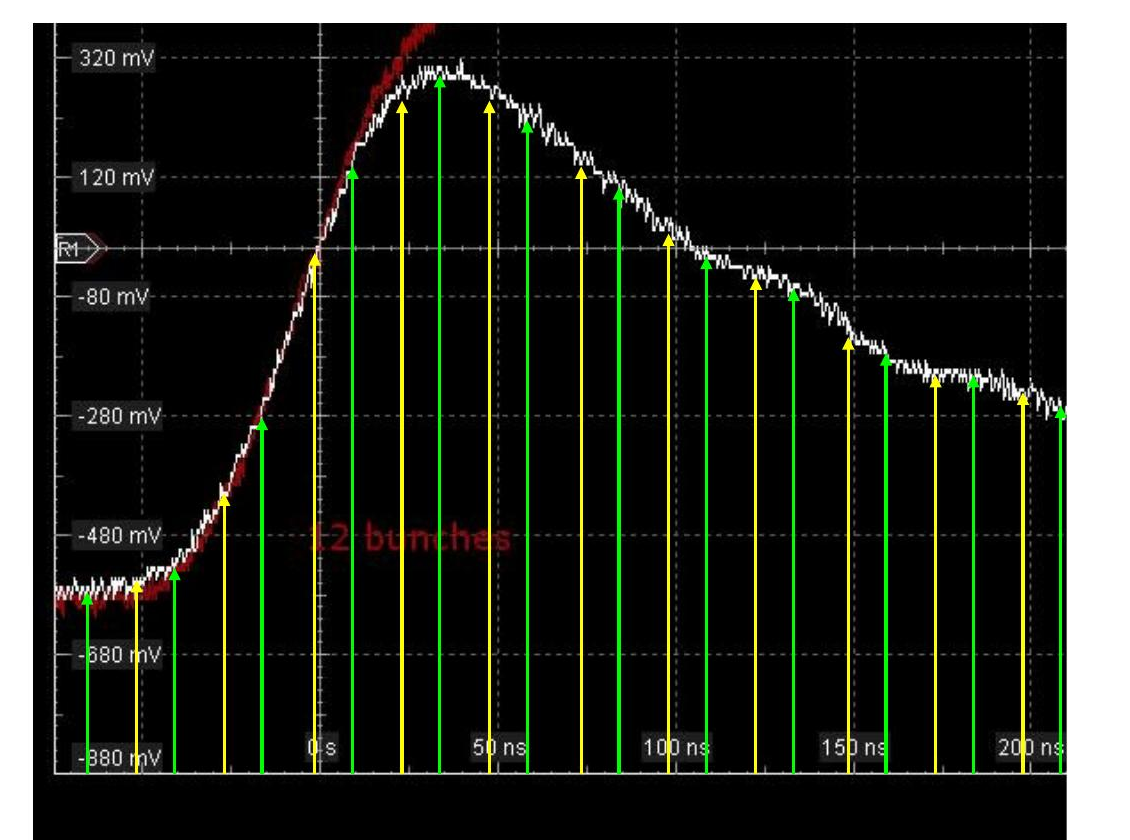}
    \caption{The same single-bunch pulse sampled at 40~Msps with two different, arbitrary sampling phases (green and yellow markers). Depending on the phase, the closest sample can fall noticeably below the true peak of the pulse.}
    \label{fig:sps_sampling_phase}
\end{figure}

Figure~\ref{fig:sps_single_bunch_model} shows a noiseless, synthesised model of the single-bunch response at the output of the~logarithmic amplifier (left) together with its magnitude spectrum (right), confirming a signal bandwidth of about 20~MHz, as expected from the bandwidth of the front-end filter. Sampling at 40~Msps is more than sufficient to satisfy the sampling theorem for a 20~MHz-wide signal (see Section~2.2.1) and therefore preserves all the information contained in the pulse; this does not mean, however, that the pulse amplitude can be read off directly from the samples. As illustrated in Fig.~\ref{fig:sps_sampling_phase}, since the ADC clock is asynchronous with respect to the beam, the sampling phase relative to the pulse is effectively random from acquisition to acquisition: depending on this phase, the samples (green or yellow markers, corresponding to two different phases) may fall well short of the true peak, leading to a potentially large error if the amplitude is simply estimated as the maximum sample value.

\begin{figure}[!ht]
    \centering
    \includegraphics[width=1.0\textwidth]{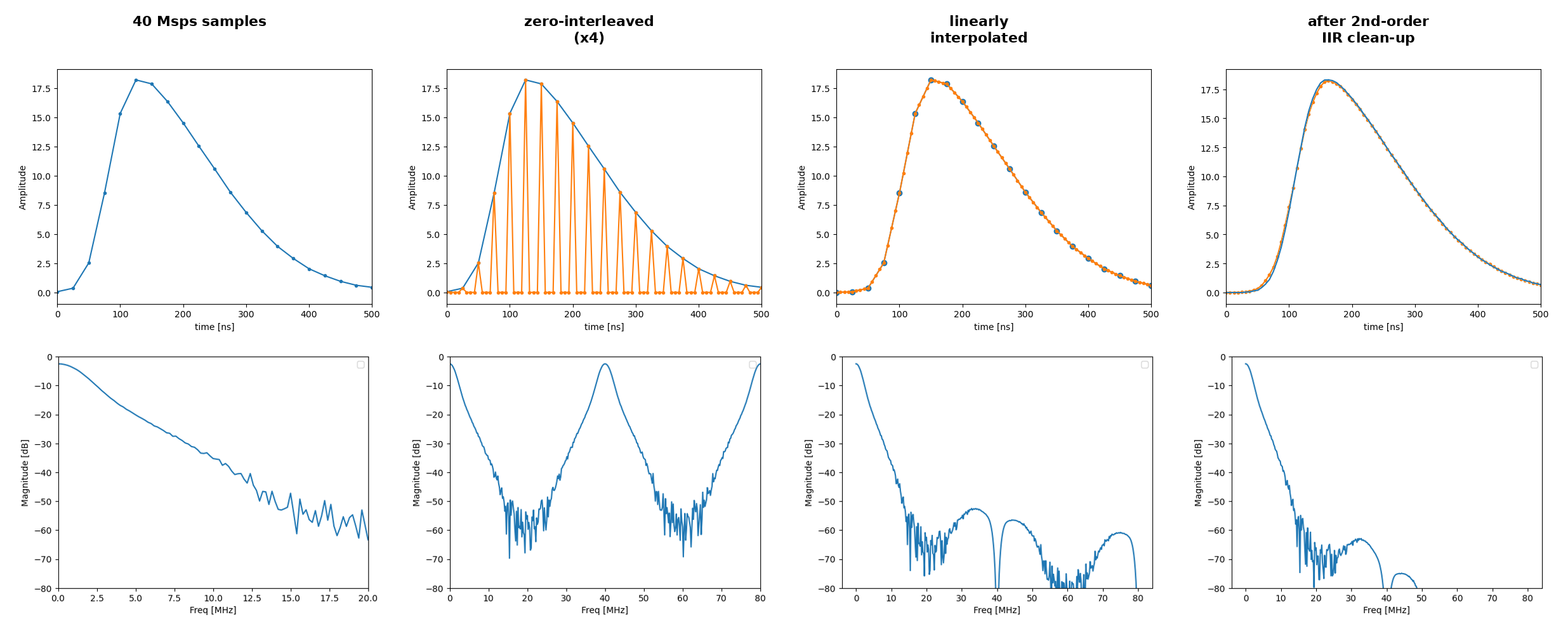}
    \caption{Digital upsampling of the single-bunch pulse by a factor of 4: the original samples at 40~Msps (left), the~zero-interleaved sequence and its aliased spectral images (second column), the result of the linear-interpolation FIR filter of Eq.~(\ref{eq:sps_interp_fir}) (third column), and the final result after the second-order IIR clean-up filter of Eq.~(\ref{eq:sps_cleanup_iir}) (right). Top row: time domain; in the first three columns, the blue line shows the original 40~Msps samples (connected for visual reference) and the orange line the signal at that processing stage, while in the last column the blue line instead shows the original, continuous analogue pulse of Fig.~\ref{fig:sps_single_bunch_model} and the orange line the final filtered result. Bottom row: magnitude spectrum of the corresponding time-domain signal at that same processing stage.}
    \label{fig:sps_upsampling}
\end{figure}

To overcome this limitation, one can exploit the fact that the full frequency-domain content of the~signal is already available in the acquired samples. The key idea is to increase the effective sampling rate digitally, through \textbf{\textit{zero interleaving}}: \(L-1\) zero-valued samples are inserted between each pair of original samples, producing a new sequence \(z[n]\) formally sampled at \(L\) times the original rate (\(L=4\) in this example). This is equivalent to sampling the original signal directly at the higher rate \(Lf_s\), and then multiplying the result by a coarser Dirac comb of period \(T_s\), which keeps only every \(L\)th sample -- the ones actually available -- and zeroes all the others. Exactly as in Section~2.2.1, multiplying by this comb convolves the spectrum with a Dirac comb of spacing \(f_s\) (Eq.~\ref{eq:sampling_replicas}), filling the enlarged spectral range -- now extending to the new Nyquist frequency \(Lf_s/2\) -- with \(L\) replicas of the original spectrum. Since no new information has actually been added, these replicas are simply aliased copies of the original spectrum, as shown in the second column of Fig.~\ref{fig:sps_upsampling}. These spectral images must be removed by low-pass filtering to obtain a correctly interpolated signal at the higher rate. Since the images to be suppressed have the same power as the wanted signal, achieving a sharp, effective separation would in principle require a~very long (ideally infinite and non-causal, i.e.\ a sinc) filter, which is impractical for real-time processing in an FPGA.

The task of the filter is considerably eased if, instead of inserting zeros, the missing samples are approximated by linear interpolation between neighbouring original samples. This is equivalent to filtering the zero-interleaved sequence \(z[n]\) with a short, triangular (Bartlett) FIR filter, of \(2L-1\) symmetric coefficients ramping linearly up to unity and back down -- and therefore, as seen in Section~4.2, with an exactly linear phase response:

\begin{equation}
y[n] = \sum_{i=-(L-1)}^{L-1}\left(1-\frac{|i|}{L}\right) z[n-i], \qquad\text{e.g., for } L=4\text{:}\quad h = \left[\tfrac{1}{4},\tfrac{1}{2},\tfrac{3}{4},1,\tfrac{3}{4},\tfrac{1}{2},\tfrac{1}{4}\right].
\label{eq:sps_interp_fir}
\end{equation}

As shown in the third column of Fig.~\ref{fig:sps_upsampling}, the resulting interpolated signal already has its spectral images significantly attenuated compared to the zero-interleaved case, even though the filter itself is extremely short. Some residual images remain, however, and a simple second-order IIR filter is enough to clean them up further:

\begin{equation}
y[n] = 0.10\,x[n] + 0.19\,x[n-1] + 0.10\,x[n-2] + 0.94\,y[n-1] - 0.33\,y[n-2].
\label{eq:sps_cleanup_iir}
\end{equation}

The result, shown in the last column of Fig.~\ref{fig:sps_upsampling}, is a signal with some residual distortion but very close to the original noiseless pulse, and from which the peak sample can be used to extract the bunch amplitude with a much smaller error than by sampling at the original rate alone.

The interpolation and clean-up filters could in principle be merged into a single, longer filter; keeping them separate, however, is more flexible in an implementation, since it allows the parameters of the IIR clean-up filter to be adjusted independently of the fixed-coefficient interpolation filter. In the real implementation, the oversampling factor actually used is 8 rather than 4, and the residual error on the extracted amplitude is completely dominated by the ADC noise rather than by any remaining uncertainty from the upsampling algorithm itself.

\subsection{Matched filter Vs. energy estimator}

\begin{figure}[!ht]
    \centering
    \includegraphics[width=0.6\textwidth]{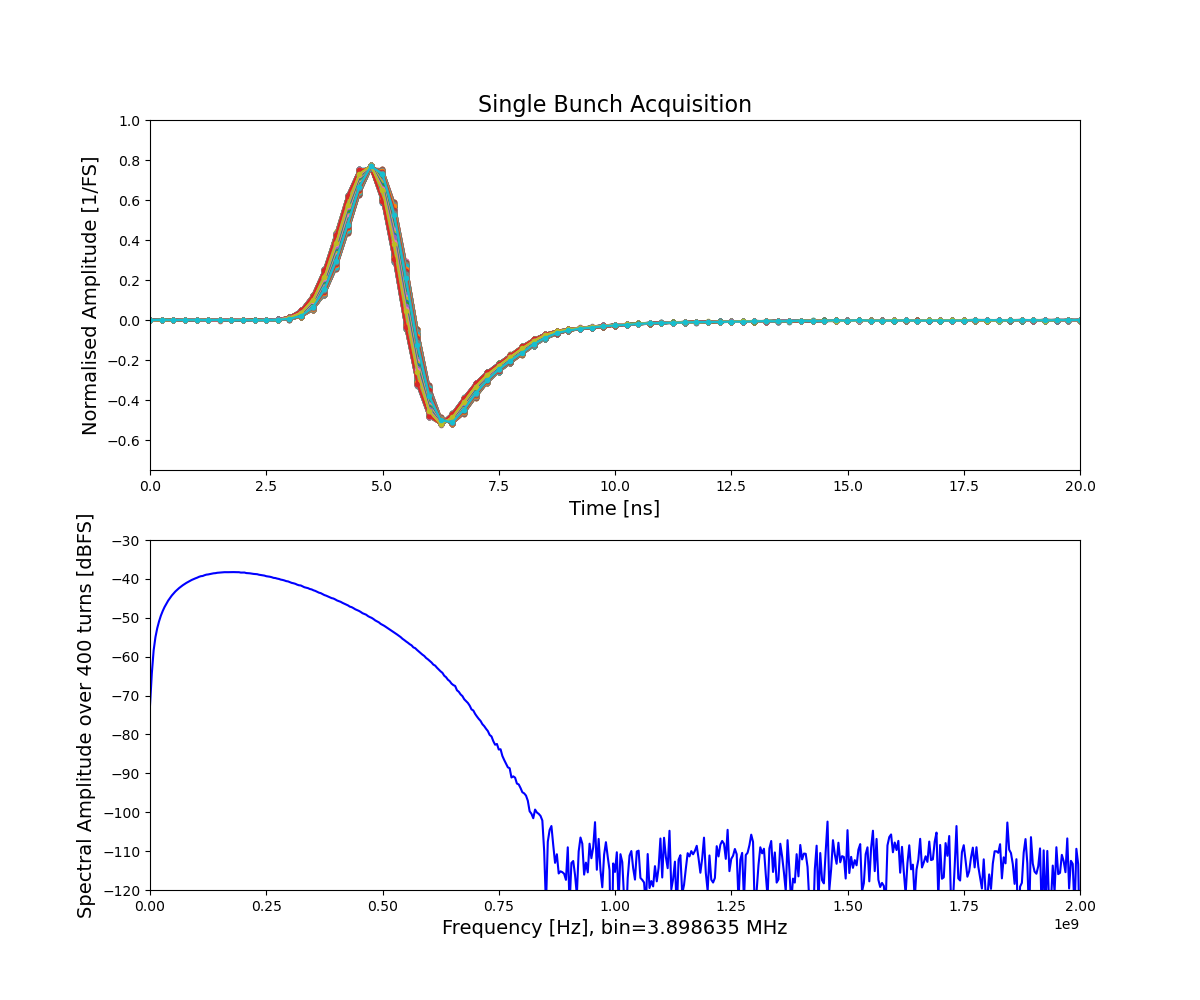}
    \caption{Top: superposition of many single-bunch pulses acquired from the same HL-LHC BPM electrode over many turns, confirming the reproducibility of the archetype pulse shape. Bottom: corresponding spectral amplitude, averaged over 400 turns.}
    \label{fig:hllhc_bpm_signal}
\end{figure}

The beam position monitors foreseen for the High-Luminosity LHC (HL-LHC) upgrade are pick-ups with four electrodes, whose relative signal amplitude, once digitised, is used to reconstruct the beam position. Each electrode produces a short analogue pulse, a few nanoseconds long, for every passing bunch, with a shape known in advance from simulation or measurement. Figure~\ref{fig:hllhc_bpm_signal} shows many such pulses, acquired from a single circulating bunch over many turns and superimposed: their remarkable reproducibility confirms that, at a given beam energy, the pulse shape can indeed be treated as a known, fixed archetype, with only its amplitude varying from acquisition to acquisition. As noted below, however, this archetype is not fixed across the whole energy ramp, since the bunch length and shape evolve as the beam is accelerated.

The HL-LHC BPM system digitises this pulse in a way that is much closer to the analogue signal than the SPS example of the previous section: rather than a resonant narrow-band filter, only a 200~MHz low-pass filter and an additional anti-aliasing filter are used, and the signal is then sampled essentially directly by a 14-bit (about 10 effective bits) ADC running at 5~Gsps. To compute the beam position, the~relative amplitude, or energy, of the signal on each of the four electrodes must be estimated as precisely as possible from these samples. Two approaches are commonly used for this task, the \textbf{\textit{matched filter}} and the \textbf{\textit{energy estimator}}, which are compared in the remainder of this section.

Let \(s[n]\) denote the known, noiseless archetype pulse shape (normalised, for convenience, to a~peak amplitude of the order of unity), so that the acquired samples can be written as
\begin{equation}
x[n] = a\, s[n] + \varepsilon[n],
\label{eq:hllhc_model}
\end{equation}
where \(a\) is the (unknown) pulse amplitude to be measured and \(\varepsilon[n]\) is additive noise, of variance \(\sigma_\varepsilon^2\) and uncorrelated from sample to sample. The \textbf{\textit{matched filter}} for \(s[n]\) is defined as the time-reversed archetype pulse, \(h[n] = s[-n]\), and, among all possible (finite-length) filters, this specific choice can be shown to maximise the signal-to-noise ratio of the filtered signal \(y[n] = x[n] \otimes h[n]\) at the point of full overlap. Indeed, for a generic filter \(h[n]\), the output at full overlap is \(a\sum_m s[m]h[-m] + \sum_m \varepsilon[m]h[-m]\), a deterministic signal term of amplitude \(a\sum_m s[m]h[-m]\) on top of a noise term of variance \(\sigma_\varepsilon^2\sum_m h^2[m]\), so that

\begin{equation}
\mathrm{SNR} = \frac{a^2\left(\sum_m s[m]h[-m]\right)^2}{\sigma_\varepsilon^2 \sum_m h^2[m]}.
\end{equation}

\begin{figure}[!ht]
    \centering
    \includegraphics[width=1.0\textwidth]{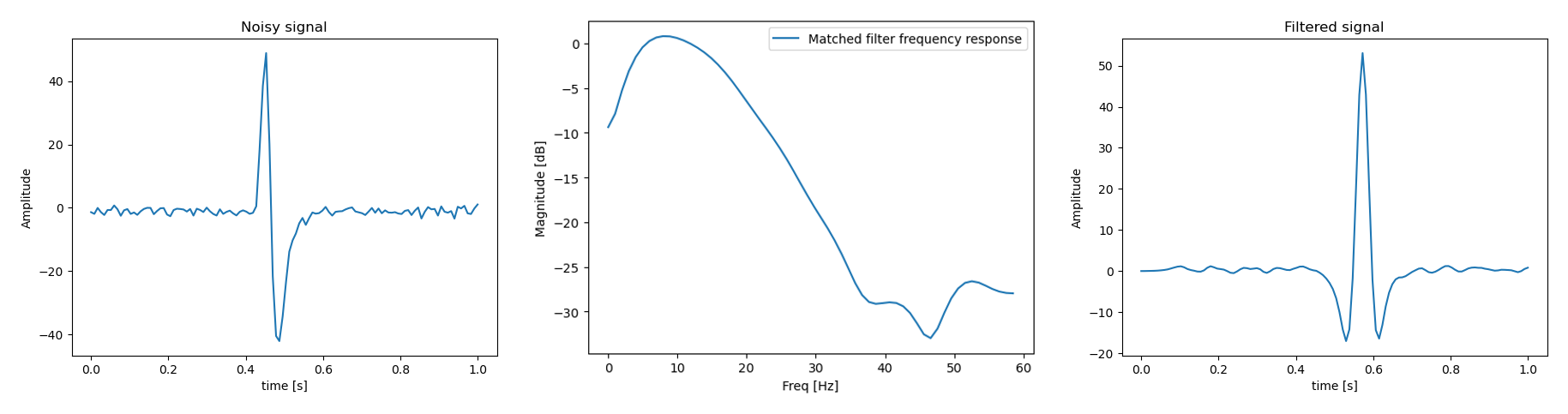}
    \caption{Illustration of the matched filter: a noisy pulse (left) is convolved with a filter matched to the archetype pulse shape, whose frequency response is shown in the centre panel; the resulting filtered signal (right) exhibits a~sharp peak at the point of full overlap, whose amplitude provides the optimal estimate of \(a\).}
    \label{fig:hllhc_matched_filter}
\end{figure}

By the Cauchy--Schwarz inequality, \(\left(\sum_m s[m]h[-m]\right)^2 \le \sum_m s^2[m] \sum_m h^2[m] = P\sum_m h^2[m]\), with equality if and only if \(h[-m]\) is proportional to \(s[m]\); substituting this bound above shows that \(\mathrm{SNR} \le a^2P/\sigma_\varepsilon^2\) for \emph{any} filter \(h[n]\), with the bound saturated precisely by the matched filter \(h[n]=s[-n]\) (see e.g.~\cite{bib:Turin}). It is important to stress that \(y[n]\) is \emph{not} meant to be a clean, restored version of the pulse itself: rather, its value at the sample of full overlap between \(x[n]\) and the archetype pulse -- i.e.\ once \(x[n]\) has slid across the entire support of \(s[n]\), which is non-zero only for \(0\le n\le N-1\) (\(N\) samples long) -- which we denote \(y[N]\), is the single point carrying this optimal estimate of the amplitude,
\begin{equation}
y[N] = a\sum_{m} s^2[m] + \sum_m \varepsilon[m]s[m] = aP + \sum_m \varepsilon[m]s[m],
\label{eq:hllhc_matched_sync}
\end{equation}
where \(P \triangleq \sum_m s^2[m]\) is the energy of the archetype pulse. This is illustrated in Fig.~\ref{fig:hllhc_matched_filter}: a noisy pulse (left) is passed through the matched filter, whose frequency response (centre) is simply that of the~archetype pulse itself; the resulting filtered signal (right) peaks sharply at the point of full overlap, and it is the amplitude of that single peak sample, not the shape of the filtered trace, that provides the~amplitude estimate.

An alternative, considerably simpler approach is the \textbf{\textit{energy estimator}}: rather than correlating the~acquired samples with a reference pulse shape, it simply sums their squares over a window of \(N\) samples spanning the pulse,
\begin{equation}
\hat{y}[n] = \sum_{i=0}^{N-1} x^2[i].
\label{eq:hllhc_energy_estimator}
\end{equation}
By Parseval's theorem, the energy estimator \(\hat{y}[n]\) is, up to a constant scaling factor, an estimate of the~energy of the sampled signal, and this estimate is independent of the sampling phase -- provided two hypotheses hold: the pulse \(s(t)\) is bandlimited within the Nyquist band of the sampling rate, so that no aliasing occurs (amply satisfied here thanks to the 200~MHz low-pass and anti-aliasing filtering ahead of the 5~Gsps ADC), and the summation window of \(N\) samples is long enough to capture the whole pulse, with negligible energy left outside it. Under these hypotheses, sampling the pulse at a different phase \(\tau\) does change the individual sample values, \(s(nT+\tau) \ne s(nT)\) in general. What is invariant is the \emph{total} energy captured by the sum: in the frequency domain, a time shift by \(\tau\) multiplies the spectrum \(S(f)\) by the all-pass factor \(e^{j2\pi f\tau}\), leaving its magnitude \(|S(f)|\), and therefore the energy \(\int|s(t)|^2\,dt\), unchanged; Parseval's theorem then guarantees that, as long as the no-aliasing hypothesis above holds, the discrete sum of squared samples equals this same energy irrespective of \(\tau\),
\begin{equation}
\sum_{n} s^2(nT+\tau) = \sum_{n} s^2(nT), \qquad \forall\,\tau.
\label{eq:hllhc_parseval_phase}
\end{equation}
This is the key property that makes the energy estimator, unlike the matched filter, completely insensitive to the sampling phase, as will be shown explicitly below. In the noiseless case, \(x[n] = a\,s[n]\), Eq.~(\ref{eq:hllhc_energy_estimator}) yields \(\hat{y}[N] = a^2\sum_i s^2[i] = a^2 P\): unlike the matched filter, whose output at the peak is \emph{linear} in the~amplitude \(a\), the energy estimator's output is \emph{quadratic} in \(a\), so that the amplitude itself is only recovered after an additional square root, \(|a| = \sqrt{\hat{y}[N]/P}\). Its great practical advantage, however, is that it requires no knowledge whatsoever of the pulse shape \(s[n]\): it can be applied identically to any signal, with no template to store or match.

Including the noise term \(\varepsilon[n]\), the two estimators can be compared directly:
\begin{equation}
y[N] = aP + \sum_{m} \varepsilon[m]s[m], \qquad
\hat{y}[N] = a^2 P + 2a\sum_{i=0}^{N-1} \varepsilon[i]s[i] + \sum_{i=0}^{N-1} \varepsilon^2[i].
\label{eq:hllhc_sync_comparison}
\end{equation}
Both estimators share a noise term proportional to the correlation of the noise with the pulse shape, \(\sum \varepsilon[i]s[i]\) -- for the energy estimator this term is simply scaled by \(2a\). The energy estimator, however, carries an additional term, \(\sum_i \varepsilon^2[i]\), coming purely from the noise power accumulated over the summation window; this term is systematically positive (a bias) and adds its own variance to the estimate, with no counterpart in the matched-filter result. For synchronous acquisition, therefore, the matched filter is generally the more accurate of the two estimators.

The situation changes, however, when the acquisition is not synchronous with the beam, as is the case for the HL-LHC BPM system, whose ADC clock runs freely with respect to the bunch arrival time. The acquired pulse is then shifted by an unknown, effectively random delay \(\tau\) with respect to the~sampling grid, \(x[n] = a\, s(nT+\tau) + \varepsilon[n]\), with \(T\) the sampling period. This time offset has no effect whatsoever on the energy estimator: as shown above, Eq.~(\ref{eq:hllhc_parseval_phase}) guarantees \(\sum_i s^2(iT+\tau) = P\) for any \(\tau\), so \(\hat{y}[N]\) retains exactly the same expression as in Eq.~(\ref{eq:hllhc_sync_comparison}). The matched filter, by contrast, now correlates the shifted, acquired pulse with a reference template built for the \emph{unshifted} pulse, which introduces a new error term:
\begin{equation}
y[N] = a\sum_{i=0}^{N-1} s(iT+\tau)s(iT) + \sum_m \varepsilon[m]s[m] = aP + a\,\delta P(\tau) + \sum_m \varepsilon[m]s[m],
\label{eq:hllhc_async}
\end{equation}
where \(\delta P(\tau)\) quantifies the mismatch between the shifted and reference pulse shapes, and depends on the unknown, effectively random sampling phase \(\tau\).

For the HL-LHC BPM system, the ADC noise \(\varepsilon[n]\) is very low, thanks to the high-resolution converter used; as a result, the mismatch term \(a\,\delta P(\tau)\) dominates over all other error contributions for the matched filter, while the energy estimator remains entirely unaffected by it. This is the main reason why the energy estimator, despite requiring an additional square root, carrying a systematic bias from the accumulated noise power, and being individually noisier in the synchronous case, is the algorithm actually selected for this system. As a further practical benefit, the energy estimator does not require the~reference pulse template to be updated as the bunch length and shape evolve during acceleration, unlike the matched filter, which would otherwise need a new template for each stage of the energy ramp.

\subsection{IQ demodulation}

Whenever the spectrum of a signal is concentrated in a few well-defined harmonics, the signal-to-noise ratio of a measurement based on that signal can be considerably improved by filtering narrowly around one of those harmonics, thereby excluding the rest of the spectrum, where mostly noise is found.

\begin{figure}[!ht]
    \centering
    \includegraphics[width=0.85\textwidth]{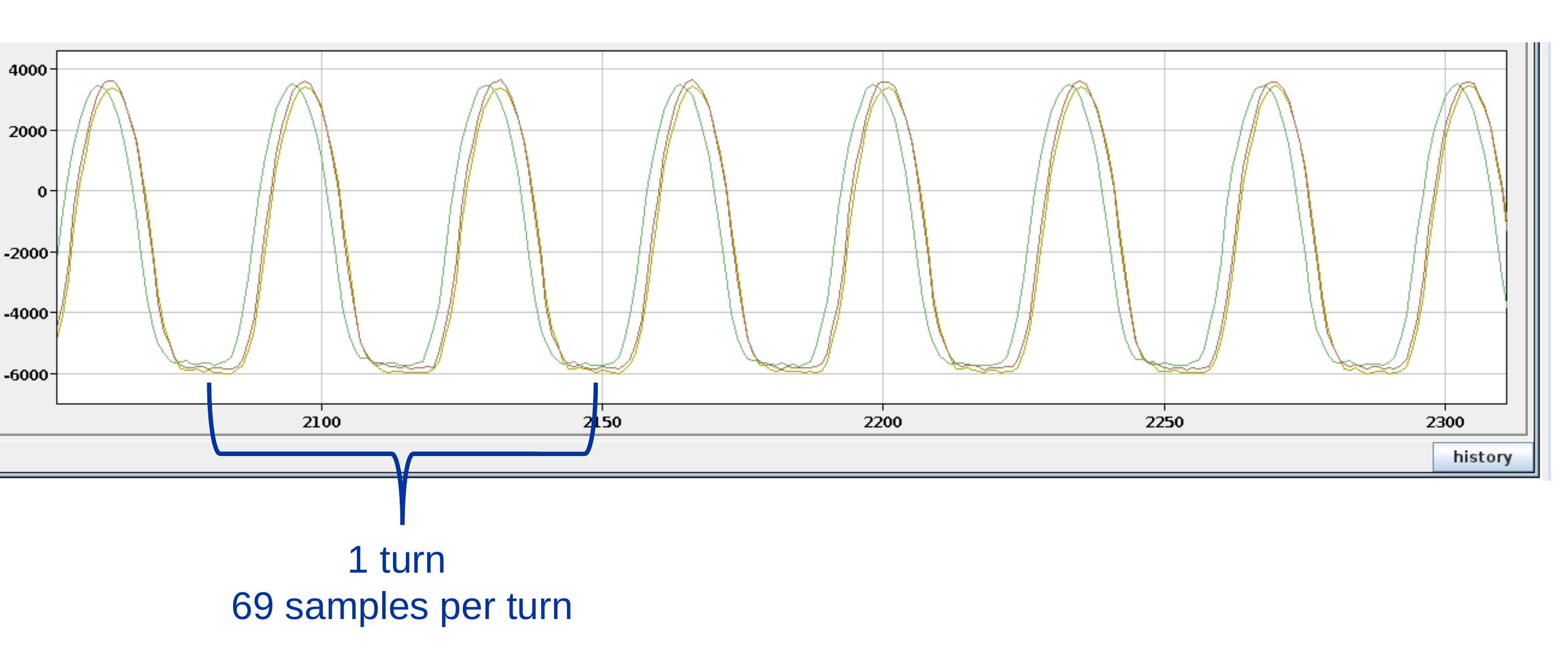}
    \caption{Real LEIR orbit pick-up signal over several consecutive turns, with the period of 69 samples per turn marked explicitly.}
    \label{fig:leir_real_acquisition}
\end{figure}

This condition is naturally satisfied whenever samples are acquired, turn after turn, from a beam circulating in a ring, as is the case for an \textbf{\textit{orbit system}} -- a system measuring the average transverse position of the beam. The signal seen by a pick-up electrode repeats almost identically from one turn to the~next, as illustrated by a real acquisition from the CERN Low Energy Ion Ring (LEIR) orbit system in Fig.~\ref{fig:leir_real_acquisition}, where the period of 69 samples per turn is marked explicitly: the signal is, to a good approximation, periodic. The trade-off of this approach is that the narrower the filter around the chosen harmonic, the lower its bandwidth, and hence the less able the resulting measurement is to track fast changes of the~orbit.

The LEIR orbit system exploits this idea, and to make it practical throughout the energy ramp of the machine, its ADC sampling frequency is locked to the accelerator's RF system: it always runs at exactly 69 times the revolution frequency, the 69\textsuperscript{th} harmonic, regardless of the beam energy. As a~result, every acquisition contains exactly 69 samples per turn (Fig.~\ref{fig:leir_real_acquisition}), and the harmonic of interest always falls at the same fixed digital frequency, a constant fraction of the sampling frequency. Had a~fixed-frequency sampling clock been used instead, the revolution-frequency changes occurring during acceleration would have appeared as a corresponding drift of the harmonic's frequency relative to the~(fixed) sampling frequency, requiring the narrow-band filter to be continuously retuned to track it.

\begin{figure}[!ht]
    \centering
    \includegraphics[width=\textwidth]{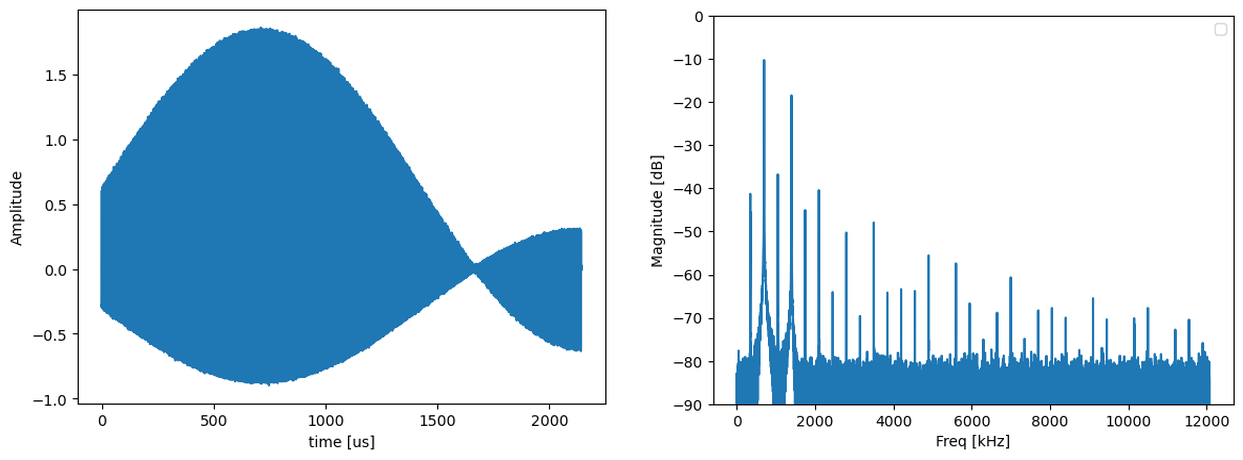}
    \caption{Left: synthetic, AC-coupled signal from a single LEIR pick-up electrode, modelled on the real LEIR pulse of Fig.~\ref{fig:leir_real_acquisition}; its amplitude variations are due to the simulated orbit trajectory, and the individual per-turn bunch pulses are not resolved at this timescale. Right: corresponding spectrum, showing the signal energy concentrated at the harmonics of the revolution frequency.}
    \label{fig:leir_synth_spectrum}
\end{figure}

To illustrate the rest of this section, a synthetic signal, modelled on the real LEIR pulse of Fig.~\ref{fig:leir_real_acquisition} but with a controlled, prescribed orbit trajectory, is used in place of a real acquisition. The signal represents the (AC-coupled) output of a single pick-up electrode over many turns: its amplitude variations are due to the simulated orbit trajectory. Its energy is nonetheless concentrated at the harmonics of the~revolution frequency, as shown in Fig.~\ref{fig:leir_synth_spectrum} (right).

\begin{figure}[!ht]
    \centering
    \includegraphics[width=\textwidth]{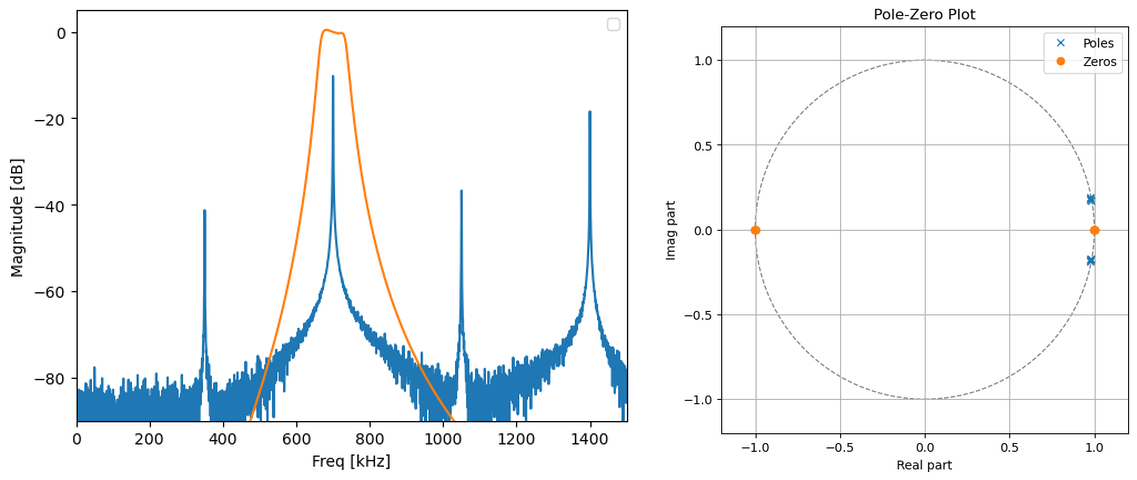}
    \caption{Left: frequency response of a narrow-band filter (orange) designed to isolate a single harmonic of the~revolution frequency (blue spectrum). Right: the corresponding pole-zero plot, showing poles very close to the~unit circle.}
    \label{fig:leir_bandpass}
\end{figure}

A first, direct approach to isolate a single harmonic would be to design a narrow-band digital band-pass filter centred on it. As shown in Fig.~\ref{fig:leir_bandpass}, however, such a filter requires poles extremely close to the unit circle, i.e.\ a high quality factor, and is therefore correspondingly sensitive to the exact target frequency and to the numerical precision of its coefficients.

The technique actually used by the LEIR orbit system, \textbf{\textit{IQ demodulation}}, avoids this difficulty by first shifting the harmonic of interest down to zero frequency (baseband), where a simple, low-order low-pass filter suffices to isolate it. Around the chosen harmonic, the acquired signal can be modelled as
\begin{equation}
x[n] = A[n]\cos(\omega_0 n + \varphi[n]) + \varepsilon[n],
\label{eq:leir_model}
\end{equation}
where \(\omega_0\) is the (fixed, thanks to the RF-locked sampling clock) digital frequency of the chosen harmonic, and \(A[n]\), \(\varphi[n]\) are its amplitude and phase, varying slowly from turn to turn and carrying the orbit information -- as noted above, \(A[n]\) is here built to follow a prescribed, synthetic orbit trajectory, for illustration purposes.

\begin{figure}[!ht]
    \centering
    \includegraphics[width=\textwidth]{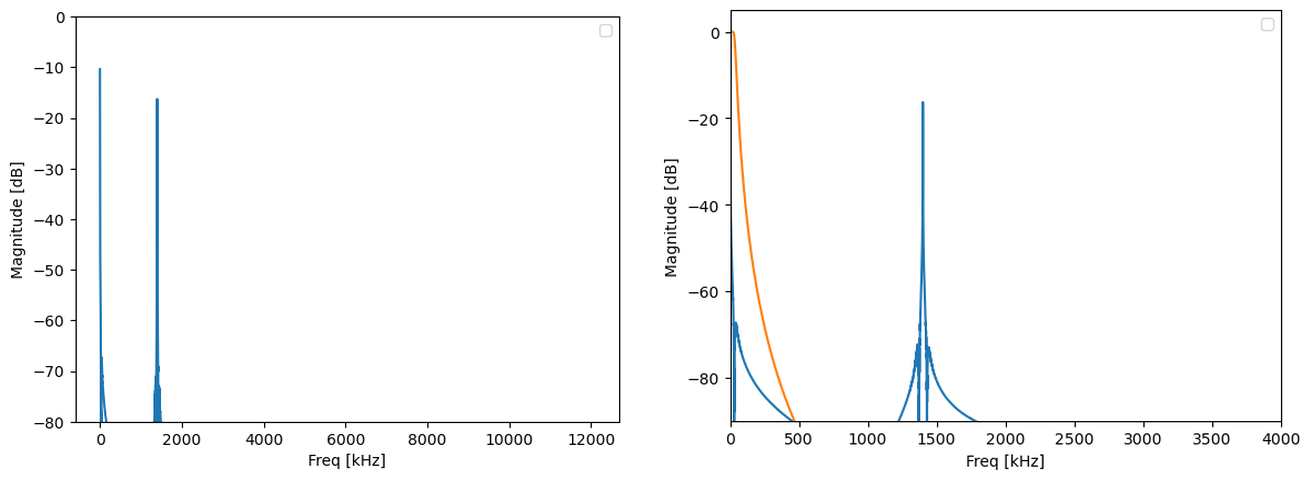}
    \caption{IQ demodulation of the LEIR orbit signal. Left: spectrum after mixing to baseband, showing the wanted term near zero frequency together with the unwanted image near \(2\omega_0\). Right: spectrum after the low-pass filter (orange), retaining only the baseband term.}
    \label{fig:leir_iq_demod}
\end{figure}

IQ demodulation multiplies \(x[n]\) by two quadrature digital local oscillators at the harmonic frequency,
\begin{equation}
I[n] = x[n]\cos(\omega_0 n), \qquad Q[n] = -x[n]\sin(\omega_0 n).
\label{eq:leir_iq}
\end{equation}
Using standard product-to-sum identities, and ignoring the noise term for clarity, Eqs.~(\ref{eq:leir_model})--(\ref{eq:leir_iq}) give
\begin{equation}
I[n] = \frac{A[n]}{2}\Big[\cos\varphi[n] + \cos(2\omega_0 n+\varphi[n])\Big], \qquad
Q[n] = \frac{A[n]}{2}\Big[\sin\varphi[n] - \sin(2\omega_0 n+\varphi[n])\Big]:
\label{eq:leir_iq_expand}
\end{equation}
each branch contains the wanted, slowly-varying baseband term together with an unwanted image term oscillating at twice the harmonic frequency, as confirmed by the spectrum of Fig.~\ref{fig:leir_iq_demod} (left). A low-pass filter, with a cut-off frequency well above the bandwidth of \(A[n]\), \(\varphi[n]\) but well below \(2\omega_0\), removes the~image term (Fig.~\ref{fig:leir_iq_demod}, right), leaving
\begin{equation}
I_{\mathrm{LP}}[n] \approx \frac{A[n]}{2}\cos\varphi[n], \qquad Q_{\mathrm{LP}}[n] \approx \frac{A[n]}{2}\sin\varphi[n],
\label{eq:leir_iq_lp}
\end{equation}
from which the amplitude and phase of the harmonic -- and, with it, the orbit information it carries -- are recovered as
\begin{equation}
A[n] = 2\sqrt{I_{\mathrm{LP}}^2[n]+Q_{\mathrm{LP}}^2[n]}, \qquad \varphi[n] = \operatorname{atan2}\big(Q_{\mathrm{LP}}[n],\,I_{\mathrm{LP}}[n]\big).
\label{eq:leir_amp_phase}
\end{equation}
Being defined through a square root, \(A[n]\) as recovered here is by construction never negative, even though the orbit itself is a signed quantity, oscillating on either side of the reference orbit. The sign is instead carried entirely by the phase \(\varphi[n]\): since \(\cos(\theta+\pi)=-\cos\theta\), a change of sign in the orbit shows up not as a negative \(A[n]\), but as \(\varphi[n]\) flipping by \(\pi\) -- for instance switching from values near \(0\) to values near \(\pi\) -- while \(A[n]\) continues to track only its magnitude.

Because the number of samples per turn is fixed once and for all by the RF-locked sampling clock, \(\cos(\omega_0 n)\) and \(\sin(\omega_0 n)\) take only 69 distinct values, repeating identically every turn: the two local oscillators can therefore be pre-computed once, as a 69-sample look-up table, and reused unchanged throughout the entire energy ramp, with no need to recompute or interpolate them as the revolution frequency evolves.

\section{Numerical representation of numbers}

While the previous section focused on the algorithms used to process a signal, this section addresses the~equally important question of how the numbers involved -- samples, coefficients, and intermediate results -- are actually represented in a digital system, and the consequences that this finite representation has on the accuracy of a measurement.

\subsection{Quantisation and quantisation error}

As introduced in Section~1, amplitude quantisation is, together with time-domain sampling, the second fundamental step in producing a digital signal. An ideal uniform quantiser \(Q(\cdot)\) maps a continuous amplitude \(x\) onto the nearest of \(2^N\) discrete levels spanning a full-scale range \(\mathrm{FSR}\), with a step, or resolution,

\begin{equation}
\Delta = \frac{\mathrm{FSR}}{2^N}.
\label{eq:quant_step}
\end{equation}

\begin{figure}[!ht]
    \centering
    \includegraphics[width=1.0\textwidth]{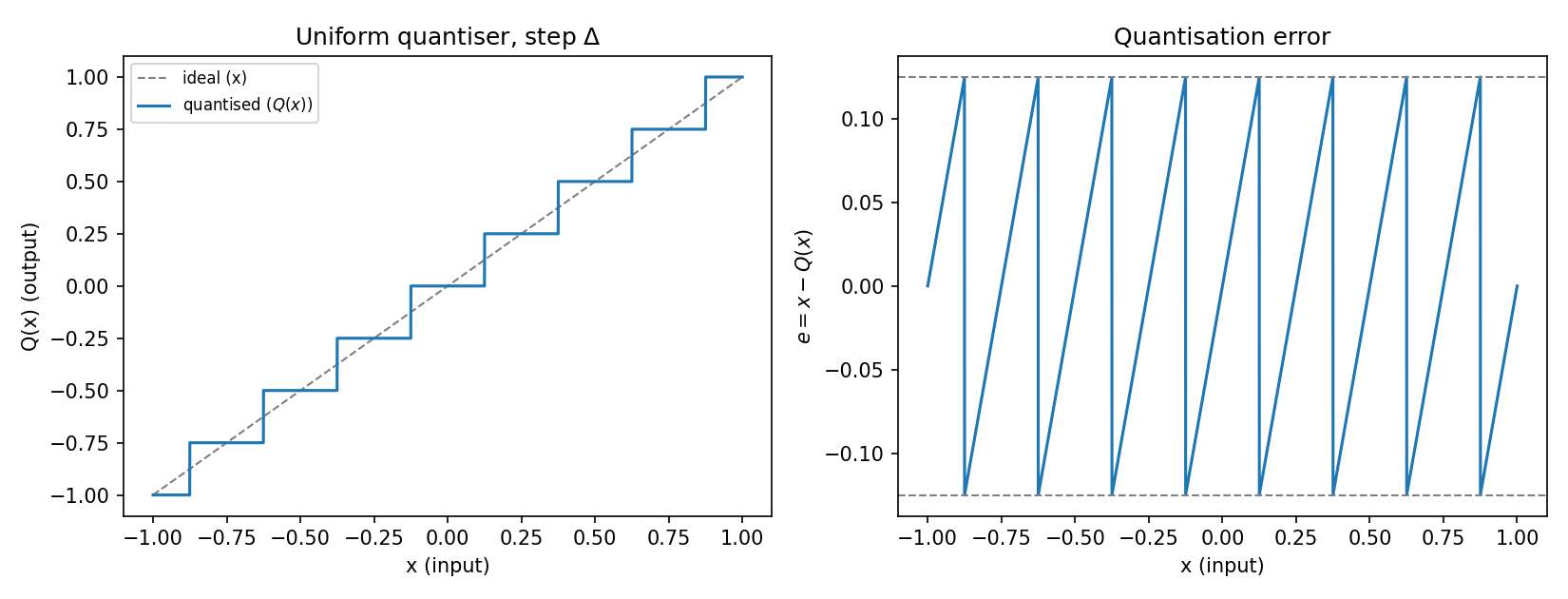}
    \caption{Transfer function of a uniform quantiser of step \(\Delta\) (left), and the corresponding quantisation error as a~function of the input (right).}
    \label{fig:quantisation_error}
\end{figure}

The quantisation error, \(e[n] = x[n] - Q(x[n])\), is illustrated in Fig.~\ref{fig:quantisation_error}. For a signal that varies by many quantisation steps from sample to sample, \(e[n]\) is well approximated by a random variable, uniformly distributed in \([-\Delta/2,\Delta/2]\) and approximately uncorrelated with the signal itself \cite{bib:OppenheimDSP,bib:Lyons}. Under this model, the quantisation error has zero mean and variance

\begin{equation}
\sigma_e^2 = \frac{1}{\Delta}\int_{-\Delta/2}^{\Delta/2} e^2\, de = \frac{\Delta^2}{12}.
\label{eq:quant_noise_variance}
\end{equation}

\subsection{Analogue to digital converters and effective number of bits}

Consider a full-scale sinusoidal input digitised by an ideal \(N\)-bit analogue-to-digital converter (ADC), \(x(t) = \frac{\mathrm{FSR}}{2}\sin(2\pi f t)\), spanning the full-scale range \(\mathrm{FSR}\). Its mean-square power is

\begin{equation}
P_{\mathrm{signal}} = \frac{1}{2}\left(\frac{\mathrm{FSR}}{2}\right)^2 = \frac{\mathrm{FSR}^2}{8}.
\end{equation}

Combining this signal power with the quantisation-noise power of Eq.~(\ref{eq:quant_noise_variance}), and using \(\mathrm{FSR}/\Delta = 2^N\) from Eq.~(\ref{eq:quant_step}), the signal-to-quantisation-noise ratio (SQNR) is

\begin{equation}
\mathrm{SQNR} = \frac{P_{\mathrm{signal}}}{\sigma_e^2} = \frac{\mathrm{FSR}^2/8}{\Delta^2/12} = \frac{3}{2}\left(\frac{\mathrm{FSR}}{\Delta}\right)^2 = \frac{3}{2}\,2^{2N}.
\end{equation}

Expressed in dB, this gives the well-known result

\begin{equation}
\mathrm{SQNR}\,[\mathrm{dB}] = 10\log_{10}\!\left(\frac{3}{2}\,2^{2N}\right) = 20N\log_{10}2 + 10\log_{10}\frac{3}{2} \approx 6.02\,N + 1.76.
\label{eq:sqnr}
\end{equation}

Real ADCs, however, are affected by additional non-idealities not included in this ideal model, such as thermal noise, differential and integral non-linearity, and aperture jitter, which reduce the actual signal-to-noise-and-distortion ratio (SINAD) achieved with a real input signal below the ideal SQNR of Eq.~(\ref{eq:sqnr}). To characterise a real converter independently of its nominal resolution, it is common practice to define the \textbf{\textit{effective number of bits}} (ENOB) as the resolution of an ideal ADC that would yield the~same SINAD:

\begin{equation}
\mathrm{ENOB} = \frac{\mathrm{SINAD}\,[\mathrm{dB}] - 1.76}{6.02}.
\end{equation}

The ENOB of a real converter typically decreases with increasing input frequency, mainly because of aperture-jitter effects, and is therefore an important figure of merit when selecting an ADC for a given beam instrumentation application.

\subsection{Integers}

\subsubsection{Unsigned integers}

The most direct digital representation of a number is the standard positional binary code. For an unsigned integer represented on \(N\) bits \(b_{N-1}b_{N-2}\ldots b_1 b_0\),

\begin{equation}
b_{N-1}b_{N-2}\ldots b_1 b_0 \ \rightarrow\ \sum_{i=0}^{N-1} b_i\, 2^i,
\end{equation}

with range \([0,\, 2^N-1]\) and resolution (step) equal to 1 -- e.g.\ the familiar 8-bit unsigned integer, UINT8. This representation extends naturally to fixed-point unsigned numbers by interpreting the last \(M\) bits as a fractional part, scaled by \(2^{-M}\) (commonly denoted, e.g., Q5.3 for \(N=8\), \(M=3\)):

\begin{equation}
b_N b_{N-1}\ldots b_1 b_0 \ \rightarrow\ \left(\sum_{i=0}^{N-1} b_i\, 2^i\right) 2^{-M},
\end{equation}

with range \([0,\, 2^{N-M} - 2^{-M}]\) and resolution \(2^{-M}\).

\subsubsection{Signed integers}

Several conventions exist to extend the above representations to signed (positive and negative) integers.

\paragraph{Sign and magnitude representation}
The most intuitive approach reserves the most significant bit as an explicit sign bit (0 for positive, 1 for negative), the remaining bits encoding the magnitude:

\begin{equation}
b_{N-1}b_{N-2}\ldots b_1 b_0 \ \rightarrow\ (-1)^{b_{N-1}} \sum_{i=0}^{N-2} b_i\, 2^i,
\end{equation}

with range \([-2^{N-1}+1,\, 2^{N-1}-1]\) and resolution 1. This representation, and the closely related \emph{one's-complement} representation (where negative numbers are obtained by bitwise complementing the~magnitude bits), suffer from the same drawback: zero has two distinct representations (\(+0\) and \(-0\)), and arithmetic circuits need to handle the sign separately from the magnitude, complicating hardware implementations.

\paragraph{Shift representation}
An alternative, sometimes referred to as \emph{offset} or \emph{excess-K} representation, simply shifts the unsigned code by half the available range:

\begin{equation}
b_{N-1}b_{N-2}\ldots b_1 b_0 \ \rightarrow\ \left(\sum_{i=0}^{N-1} b_i\, 2^i\right) - 2^{N-1},
\end{equation}

with range \([-2^{N-1},\, 2^{N-1}-1]\) and resolution 1. This representation has a single code for zero and increases monotonically with the binary code, which makes it convenient, for instance, to represent the biased exponent of a floating-point number, discussed below.

\paragraph{Two's complement representation}
By far the most widely used representation in modern digital hardware is the \textbf{\textit{two's complement}} representation:

\begin{equation}
b_{N-1}b_{N-2}\ldots b_1 b_0 \ \rightarrow\ \left(\sum_{i=0}^{N-2} b_i\, 2^i\right) - b_{N-1}\, 2^{N-1},
\end{equation}

with range \([-2^{N-1},\, 2^{N-1}-1]\) and resolution 1. It has a single representation of zero, and, crucially -- unlike any of the three representations described above, all of which require separate handling of the sign or a different addition rule for positive and negative operands -- addition of two's complement numbers can be performed with exactly the same hardware adder used for unsigned numbers, simply working modulo \(2^N\): the \(2^N\) available codes can be thought of as arranged on a circle, with addition corresponding to a step along the circle, and overflow corresponding to wrapping around the boundary between the most-positive and the most-negative codes.

The representation takes its name from the operation used to negate a number. To form \(-x\) from a~positive integer \(x\), one \emph{complements} \(x\) with respect to \(2^N\), i.e.\ computes \(2^N-x\) (working modulo \(2^N\)). In binary, this is most conveniently carried out in two steps: first bitwise-complementing every bit of \(x\) (flipping each 0 to 1 and vice versa), which yields the \emph{one's complement} of \(x\), equal to \(2^N-1-x\); then adding 1, to reach \(2^N-x\). This ``complement, then add one'' recipe is the standard algorithm used in practice to negate a two's complement number, and gives the~representation its name: it is the~complement with respect to the radix, \(2^N\) (``two's'' complement), as opposed to the one's-complement representation mentioned above, which stops at the first step, complementing with respect to \(2^N-1\) rather than \(2^N\), and is left with the two representations of zero noted there.

This construction, together with a straightforward extension to a fixed-point format identical to the~one introduced for unsigned integers, explains why two's complement is the standard choice for signed integer and fixed-point arithmetic in essentially all modern processors, DSPs and FPGAs.

\subsection{Real numbers}

\subsubsection{The issue with real numbers in the digital domain}

Real numbers can, in general, require infinite precision and therefore cannot be represented exactly with a finite number of bits. Two main strategies are used in practice to represent real numbers digitally: \emph{fixed-point} representation, a direct extension of the integer representations discussed above, and \emph{floating-point} representation, which trades a constant absolute resolution for a much wider dynamic range.

\subsubsection{Fixed-point representation}

As already introduced above, a fixed-point number is simply a signed (or unsigned) integer in which the~position of the binary point is conventionally fixed at \(M\) bits from the right, giving a constant resolution of \(2^{-M}\) over the whole representable range. Fixed-point arithmetic reuses standard integer arithmetic units and is therefore very efficient to implement in hardware, particularly on FPGAs and DSP processors; its main limitation is the restricted dynamic range set by the fixed number of integer bits, which can lead to overflow, or, conversely, to a poor relative resolution for signals with a wide dynamic range \cite{bib:Lyons}.

\subsubsection{Floating-point representation}

A \textbf{\textit{floating-point}} number instead represents a real value as a signed, normalised significand multiplied by a power of two, encoded with \(N = F+E+1\) bits, split into a sign bit, \(E\) bits of (shift-represented) exponent, and \(F\) bits for the fractional part of the significand:

\begin{equation}
b_{N-1}b_{N-2}\ldots b_1 b_0 \ \rightarrow\ (-1)^{b_{N-1}} \left(1 + 2^{-F}\sum_{i=0}^{F-1} b_i\, 2^i\right) 2^{\left(\sum_{i=0}^{E-1} b_{i+F}\, 2^i\right) - 2^{E-1}},
\end{equation}

i.e.\ \(\mathrm{value} = \mathrm{significand} \cdot 2^{\mathrm{exponent}}\), with

\begin{equation}
\mathrm{significand} \in [1,\, 2-2^{-F}] \cup [-1,\, -(2-2^{-F})], \qquad \mathrm{exponent} \in [-2^{E-1},\, 2^{E-1}-1].
\end{equation}

The significand is normalised with an implicit leading 1, so that only the fractional bits need to be stored, and the exponent uses a shift representation so that its binary code increases monotonically with its value. A crucial consequence of this construction is that the resolution between two adjacent representable numbers, \(2^{\mathrm{exponent}-F}\), is \emph{not constant}, but scales with the magnitude of the number itself: floating-point numbers therefore provide an approximately constant \emph{relative} precision over a very large dynamic range, in contrast with the constant \emph{absolute} precision, but limited range, of fixed-point numbers. Zero is a special case that cannot be represented by the normalised form above, and is conventionally encoded as an exception, replacing the two codes of smallest modulus (allowing both a \(+0\) and a \(-0\)). The construction just described is, in essence, exactly the one specified by the widely adopted IEEE~754 standard for single- and double-precision floating-point arithmetic \cite{bib:IEEE754}, used throughout general-purpose computing and, increasingly, in FPGA- and DSP-based processing chains.

\section{Hardware solutions for DSP implementation}

Once a DSP algorithm has been designed, it must be mapped onto an actual computing platform. Four broad families of devices are commonly used to implement digital signal processing algorithms in beam instrumentation systems: the central processing unit (CPU), the graphics processing unit (GPU), the~digital signal processor (DSP), and the field-programmable gate array (FPGA). Each family represents a~different trade-off between flexibility, parallelism, latency, power efficiency and development effort, and the~choice of platform is ultimately driven by the specific requirements of the application.

\begin{itemize}
\item \textbf{CPU} -- a general-purpose processor optimised for versatility, low-latency branching and single-thread performance, well suited to complex logic, operating systems and varied, sequential or decision-heavy tasks. Its main limitation for DSP applications is a comparatively limited degree of parallelism.
\item \textbf{GPU} -- a massively parallel processor with thousands of small cores designed for SIMD/SIMT execution, originally developed for graphics and now widely used for machine learning and high-performance computing. It offers very high throughput for vectorisable, matrix-oriented workloads, at the price of higher latency, poor efficiency for irregular control flow, and high power consumption.
\item \textbf{DSP} -- a microprocessor specialised for fast arithmetic on sampled data (multiply-accumulate operations, filtering, FFTs), typically featuring a dedicated instruction set, hardware multiply-accumulate units and circular buffers. It provides efficient, deterministic, low-power real-time signal processing, at the cost of a more limited degree of parallelism than GPUs or FPGAs.
\item \textbf{FPGA} -- a reconfigurable hardware device built from logic blocks, look-up tables, dedicated DSP slices and programmable interconnects, in which the hardware architecture itself is described using a hardware description language (e.g.\ VHDL or Verilog). FPGAs offer true hardware-level parallelism, deterministic and very low latency, and arbitrary fixed-point precision, making them particularly well suited to high-throughput, real-time streaming and control-loop applications; their main drawbacks are a comparatively difficult and time-consuming design process and slower clock speeds than CPUs or GPUs.
\end{itemize}

In short: an FPGA provides reconfigurable hardware for ultra-parallel, low-latency, deterministic workloads; a DSP is a processor optimised for fast, real-time signal arithmetic; a CPU is a general-purpose processor optimised for versatility and low-latency control; and a GPU is a massively parallel processor optimised for large-scale numeric throughput. A qualitative comparison of the four families is summarised in Table~\ref{tab:hw_comparison}.

\begin{table}[!ht]
\centering
\caption{Qualitative comparison of the four hardware families commonly used for DSP implementation.}
\label{tab:hw_comparison}
\resizebox{\textwidth}{!}{%
\begin{tabular}{lllll}
\hline
\textbf{Feature} & \textbf{FPGA} & \textbf{DSP} & \textbf{CPU} & \textbf{GPU} \\
\hline
Parallelism & Hardware-level, true parallel & Moderate (MAC units) & Low--moderate (few cores) & Very high (thousands of threads) \\
Latency & Very low, deterministic & Low, often deterministic & Low for control, workload-dependent & Higher, less deterministic \\
Throughput & High for pipelines & Medium & Medium & Very high \\
Best precision & Arbitrary fixed-point & Fixed- and floating-point & Float or integer & Float or integer \\
Programming difficulty & Hard (HDL, toolchains) & Medium & Easy & Medium (CUDA, OpenCL) \\
Power efficiency & Excellent (custom design) & Good & Moderate & Worst (high power) \\
Real-time behaviour & Excellent & Excellent & Good & Poor \\
Cost & Medium--high & Low--medium & Low--medium & Medium--high \\
\hline
\end{tabular}%
}
\end{table}

\subsection{Blurring the borders: RF system-on-chip technology}

A recent trend in high-speed instrumentation is the increasing integration of the four hardware families discussed above onto a single device. \textbf{\textit{RF system-on-chip (RF~SoC)}} devices combine, on a single die, a~processing system based on general-purpose ARM CPU cores (e.g.\ a quad-core Cortex-A53 application processor and a dual-core Cortex-R5 real-time processor), programmable logic equivalent to a modern FPGA fabric with embedded DSP blocks, and, most notably, directly integrated RF-class analogue-to-digital and digital-to-analogue converters, with sampling rates of several Gsps for the ADCs and up to about 10~Gsps for the DACs \cite{bib:RFSoC}. By bringing high-speed data conversion directly onto the same die as the processing logic, RF SoC devices considerably simplify the analogue front-end of high-bandwidth beam instrumentation systems, and increasingly blur the traditional boundaries between the CPU, DSP and FPGA hardware families discussed above.

\section{Conclusion}

These proceedings have summarised a lecture on digital signal processing given at the CERN Accelerator School on Beam Instrumentation. Starting from the basic notion of a signal and its digitisation, the~discussion first revisited frequency-domain analysis, from the Fourier series to the Fourier transform, and showed how time-domain sampling and windowing lead naturally from the continuous Fourier transform to the DTFT and, finally, to the DFT implemented in software and hardware.

Building on this frequency-domain foundation, filters were introduced first in the analogue domain, through the differential equation of an LTI system and the resulting decomposition of its Bode response into individual pole and zero contributions, and then in the digital domain, through the analogous difference equation and its associated Z-transform. The two main digital filter families, FIR and IIR, were then examined in turn: FIR filters, illustrated with a notch filter cancelling a narrow-band disturbance in an LHC tune measurement and a moving-average low-pass filter; and IIR filters, illustrated with the LHC orbit filter, the geometric interpretation of stability in terms of pole location, and the bilinear transformation used to design digital IIR filters from well-established analogue prototypes. Together, these examples illustrate how these concepts translate directly into solutions for common beam instrumentation problems.

These building blocks were then combined in three complete, real-world case studies. The single-bunch position measurement in the CERN SPS showed how a resonant analogue front-end, logarithmic compression and asynchronous sampling require a digital upsampling stage -- combining a short linear-interpolation FIR filter with a second-order IIR clean-up filter -- to recover an accurate amplitude from a~single sample per acquisition. The comparison between the matched filter and the energy estimator for the HL-LHC BPM system then illustrated how the best algorithm on paper is not always the best choice in practice: although the matched filter is the optimal estimator for synchronous acquisition, the energy estimator's insensitivity to the sampling phase, guaranteed by Parseval's theorem, makes it the preferred choice for a system whose digitisation is not synchronous with the beam. Finally, IQ demodulation in the CERN LEIR orbit system showed how locking the sampling clock to the machine's RF system keeps a~chosen revolution-frequency harmonic at a fixed digital frequency throughout the energy ramp, and how shifting that harmonic down to baseband turns an otherwise impractical, high-quality-factor band-pass filtering problem into a simple low-pass one.

Finally, the discussion turned to the numerical representation of the samples and coefficients themselves, from quantisation and the effective number of bits of an ADC, through the various integer representations, to fixed- and floating-point real numbers, before closing with an overview of the CPU, GPU, DSP and FPGA hardware families used to implement DSP algorithms in practice, and the increasing integration of these functions in modern RF system-on-chip devices.

Digital signal processing is a vast field, and this contribution could only cover a limited, application-driven subset of it. The interested reader is encouraged to consult the references below \cite{bib:OppenheimSS,bib:OppenheimDSP,bib:Lyons} for a more complete and rigorous treatment of the topics introduced here.

\end{document}